\documentclass[10pt]{iopart}
\usepackage{graphicx}
\usepackage{bm}
\usepackage{amssymb}
\usepackage{mathrsfs}
\usepackage{amsfonts}
\usepackage[left=2.5cm,top=4.5cm,right=2.5cm,bottom=3cm]{geometry}
\usepackage{setspace}
\usepackage{cite}
\DeclareMathSizes{10.95}{10}{7}{7}
\def \textbf{\bi}

\newtheorem{theorem}{{\bf Theorem}}

\newtheorem{lemma}{{\bf Lemma}}

\newtheorem{remark}{{\bf Remark}}

\newcommand{\qed}{\nobreak \ifvmode \relax \else
      \ifdim\lastskip<1.5em \hskip-\lastskip
      \hskip1.5em plus0em minus0.5em \fi \nobreak
      \vrule height0.75em width0.5em depth0.25em\fi}

\begin{document}

\title[Hamiltonian Formalism of Extended Magnetohydrodynamics]{Hamiltonian Formalism of Extended Magnetohydrodynamics}

\author{H. M. Abdelhamid$^{1,2}$, Y. Kawazura$^1$\& Z. Yoshida$^1$}

\address{$^1$Graduate School of Frontier Sciences, University of Tokyo, Kashiwanoha, Kashiwa, Chiba 277-8561, Japan}
\address{$^2$Physics Department, Faculty of Science, Mansoura University, Mansoura 35516, Egypt}
\ead{hamdi@ppl.k.u-tokyo.ac.jp}
\vspace{10pt}
\begin{indented}
\item[]January 2014
\end{indented}

\begin{abstract}
The extended magnetohydrodynamics (MHD) system, including the Hall effect and the electron inertia effect,
has a Hamiltonian structure embodied by a noncanonical Poisson algebra on an infinite-dimensional phase space.
A nontrivial part of the formulation is the proof of Jacobi's identity for the Poisson bracket.
We unearth a basic Lie algebra that generates the Poisson bracket.
A class of similar Poisson algebra may be generated by the same Lie algebra,
which encompasses the Hall MHD system and inertial MHD system.
\end{abstract}

%
%
%
%
%

\section{Introduction}
Hamiltonian formalisms help us to elucidate geometrical structures of evolving systems.
For example, non-trivial center of the governing Poisson algebra 
(which makes the Hamiltonian system \emph{noncanonical}\,\cite{Morrison98})
yields Casimir invariants and foliates the phase space;
the corresponding topological constraints give rise to interesting equilibrium points on leaves,
forbid various instabilities, 
or separate different regimes creating hierarchical structure in the phase space\,\cite{YoshidaMorrison14}. 

The Hamiltonian formalism of the ideal magnetohydrodynamics (MHD) system was given for the first time by Morrison and Green\,\cite{MorrisonR80, MorrisonR82E};
see \cite{Morrison82,Hameiri} for recent studies on the noncanonical properties of the Poisson bracket.
The ideal MHD model often falls short of describing 
interesting phenomena in plasmas originating from different scale hierarchies
which are scaled by ion and electron inertial lengths.
For example, the electric field in the direction of the magnetic field must vanish in
the ideal MHD model, by which the topology of magnetic field lines (such as the linking numbers or
writhe) are invariant
\footnote{The magnetic helicity is a Casimir invariant pertinent to this topological constraint.
However, the constraint is much stronger than that depicted by the magnetic helicity;
see \cite{YoshidaMorrison14} for the delineation of the local topological constraints by
constructing hierarchical phase spaces.
}
In a high-temperature (collisionless) plasma,
topological change of magnetic field lines can occur in a small scale on which the electron inertia produces a finite
parallel electric field, which, however, is ignored in the ideal MHD model. Many different attempts have been made to generalize the model to include small-scale effects,
and formulate them as Hamiltonian systems;
see \cite{Holm87} for the Hamiltonian form of Hall MHD,
\cite{Kawazura2012} for the Casimir invariants of noncanonical Hall MHD,
\cite{YoshidaHameiri13} for the canonized Hamiltonian formalism of Hall MHD and its action principle
delineating the limiting path to the ideal MHD system.
Another important effect is due to the electron inertia,
which brings about a finite parallel electric field, allowing magnetic field lines reconnect.
In this direction, two-dimensional models have been intensively studied; see
\cite{Schep94, Pegoraro98, Tassi8, Tassi9, Tassi10}.



The aim of this work is to formulate a Hamiltonian system of an extended MHD model
which subsumes ideal MHD, Hall MHD, as well as inertial MHD models;
the small-scale effects are scaled by factors representing the ion skin depth and electron skin depth.
The key issue of the modeling is the reduction from the ion-electron two-fluid model by imposing the charge neutrality condition;
the two-fluid model has a natural Hamiltonian formalism (for example; see \cite{YoshidaMahajan12}), 
which, however, is not apparent in models assuming the charge neutrality condition.
In section 2, we explain the derivation of an extended MHD model from the two-fluid model. 
In section 3, we formulate the Hamiltonian and Poisson operator.
Section 4 is devoted for the proof of Jacobi's identity required for the Poisson bracket.
In section 5, we conclude this paper with some remarks.

\section{Extended MHD model}

\indent \indent Reducing the two-fluid model of plasmas by quasineutrality condition,
we obtain a system of equations governing the total mass density
$\rho=\left(m_{i}n_{i}+m_{e}n_{e}\right)\approx m_{i}n_{i}$ ($m_i$ is the ion mass, $m_e ~(\ll m_i)$ is the electron mass,
$n_i$ is the ion number density, and $n_e$ is the electron number density;
by charge neutrality, we put $n_{i}=n_{e}=n$),
the flow velocity $\textbf{V}=\left(n_{i}m_{i}\textbf{V}_{i}+n_{e}m_{e}\textbf{V}_{e}\right)/ \rho\approx\textbf{V}_{i}$,
and the magnetic field $\textbf{B}$.
By adding of the electron and ion continuity equations, we obtain the mass conservation law
\begin{eqnarray}
\frac{\partial \rho}{\partial t}=-\nabla\cdot\left(\rho\textbf{V}\right).
\end{eqnarray}
Summing the equations of motion of both components, yields the momentum equation
\begin{eqnarray}
\rho\left(\frac{\partial\textbf{V}}{\partial t}  + \left(\textbf{V}\cdot\nabla\right)\textbf{V}\right) = -\nabla p+\textbf{J}\times\textbf{B}-\frac{m_{e}}{e}\left(\textbf{J}\cdot\nabla\right)\frac{\textbf{J}}{en},
\end{eqnarray}
where $p=p_{i}+p_{e}$ is the pressure($p_{i}$ is the ion pressure, and $p_{e}$ is the electron pressure), $e$ is the charge, and $\textbf{J}$ is the current density. 
\begin{eqnarray}
\label{3}
\textbf{J}=en\left(\textbf{V}-\textbf{V}_{e}\right) .
\end{eqnarray}
Substituting of equation (\ref{3}) in the electron equation of motion we obtain the fluid equation of the current density
\small
\begin{eqnarray}
\frac{m_{e}}{e^{2}}\left[ \frac{\partial}{\partial t}\left(\frac{\textbf{J}}{n}\right)+(\textbf{V}\cdot\nabla)\frac{\textbf{J}}{n}+\left(\frac{\textbf{J}}{n}\cdot\nabla\right)\textbf{V}-\left(\frac{\textbf{J}}{n}\cdot\nabla\right)\frac{\textbf{J}}{n}\right]-\frac{1}{en}\nabla p_{e}= \left[\textbf{E}+\textbf{V}\times\textbf{B}-\frac{1}{en} \textbf{J}\times\textbf{B}\right]\nonumber\\
\end{eqnarray}
\normalsize
where $\mathbf{E}$ is the electric field. Summarizing the equations and normalizing variables in the standard Alfv{\'e}n units, we obtain a system of governing equations
\begin{eqnarray}
\frac{\partial \rho}{\partial t}=-\nabla\cdot\left(\rho\textbf{V}\right),
\end{eqnarray}
\begin{eqnarray}
\rho\left(\frac{\partial\textbf{V}}{\partial t}  + \left(\textbf{V}.\nabla\right)\textbf{V}\right) = -\nabla p+\textbf{J}\times\textbf{B}-d^{2}_{e}\left(\textbf{J}\cdot\nabla\right)\frac{\textbf{J}}{n},
\end{eqnarray}
\begin{eqnarray}
\label{7}
\textbf{E}+\textbf{V}\times\textbf{B}= -\frac{d_{i}}{n}\nabla p_{e}+d_{i}\frac{\textbf{J}}{n}\times\textbf{B}&+d^{2}_{e}\left[ \frac{\partial}{\partial t}\left(\frac{\textbf{J}}{n}\right)+(\textbf{V}\cdot\nabla)\frac{\textbf{J}}{n}+\left(\frac{\textbf{J}}{n}\cdot\nabla\right)\textbf{V}\right]\nonumber\\&-d_{i}d^{2}_{e}\left(\frac{\textbf{J}}{n}\cdot\nabla\right)\frac{\textbf{J}}{n},
\end{eqnarray}
where $d_{e}=c/(\omega_{pe}L)$ is the normalized electron skin depth, $d_{i}=c/(\omega_{pi}L)$ is the normalized ion skin depth, $\omega_{pe}$ and $\omega_{pi}$ are the electron and ion plasma frequencies($L$ is the system size).

The above equation are coupled with the pre-Maxwell equations
\begin{eqnarray}
\label{8}
\nabla\times\textbf{E}=-\frac{\partial\textbf{B}}{\partial t}\qquad   \textrm{and}\qquad  \nabla\times\textbf{\textbf{B}}=\textbf{J}.
\end{eqnarray}
\section{Hamiltonian formalism of extended MHD}
\subsection{ Noncanonical Hamiltonian systems} 
\indent \indent We prepare general notations to formulate Hamiltonian systems,
a general Hamiltonian's equation can be written as
\begin{eqnarray} 
\label{9}
\partial_t u= \mathcal{J}(u)\partial_u\mathscr{H}(u),
\end{eqnarray}
where the state vector $u$ is a point in a phase space (Hilbert space) $X$,
$\mathscr{H}\left(u\right)$ is a Hamiltonian (a smooth function on $X$),
$\partial_u \mathscr{H}$ is the gradient of $\mathscr{H}$ in $X$, and
$\mathcal{J}\left(u\right)$ is a Poisson operator (co-symplectic 2-covector).
In the latter discussions, the state vector $u$ is a vector-valued function on a base space $\Omega\subset R^3$. 
The inner product of the phase space $X$ is defined by $\langle u, v \rangle = \int_\Omega u\cdot v d^3x$.

We define a bilinear form
\begin{eqnarray} 
\{F,G\}  
=\left\langle \partial_u F,\mathcal{J}\partial_u G \right\rangle ,
\end{eqnarray}
where $F$ and $G$ are scalar smooth functionals on the phase space $X$.
For $\{~,~\}$ to be a Poisson bracket, we demand
(1) antisymmetry  $\{F,G\} +\{G,F\}=0$,
(2) Jacobi's identity $\{E,\{F,G\}\}+\{G,\{E,F\}\}+\{F,\{G,E\}\}=0$,
and (3) Leibniz property $\{FG,E\} = F\{G,E\} + G\{F,E\}$.  

The adjoint representation of Hamilton's equation (\ref{9}) reads, for an arbitrary observable $F\in C^\infty_{\{~,~\}}\left(X\right)$.
\begin{eqnarray}
\label{11}
\frac{d}{dt}F= \{ F,\mathscr{H}\}.
\end{eqnarray}

\subsection{Poisson algebra of extended MHD}
\indent \indent Operating $\nabla\times$ on both side of (\ref{7}), assuming barotropic pressures ($\rho^{-1}\nabla p=\nabla h\left(\rho\right)$, $\rho^{-1}\nabla p_{e}=\nabla h_{e}\left(\rho\right)$, where $h\left(\rho\right)$ is the total enthalpy and $h_{e}\left(\rho\right)$ is the electron enthalpy) and using (\ref{8}), we obtain a system of evaluation equations 
\begin{eqnarray}
\label{12}
\frac{\partial \rho}{\partial t}=-\nabla\cdot\left(\rho\textbf{V}\right),
\end{eqnarray}
\begin{eqnarray}
\label{13}
\frac{\partial \textbf{V}}{\partial t}=-\left(\nabla\times\textbf{V}\right)\times\textbf{V}-\nabla\left(h+\frac{V^{2}}{2}\right)+\rho^{-1} \left(\nabla\times\textbf{B}\right)\times\textbf{B}^{\ast}-d^{2}_{e}\nabla\left(\frac{\left(\nabla\times\textbf{B}\right)^{2}}{2\rho^{2}}\right),
\end{eqnarray}

\begin{eqnarray}
\label{14}
\frac{\partial \textbf{B}^{\ast}}{\partial t}=\nabla\times\left(\textbf{V}\times\textbf{B}^{\ast}\right)- d_{i} \nabla\times\left(\rho^{-1} \left(\nabla\times\textbf{B}\right)\times\textbf{B}^{\ast}\right)+d^{2}_{e} \nabla\times\left(\rho^{-1} \left(\nabla\times\textbf{B}\right)\times\left(\nabla\times\textbf{V}\right)\right),\nonumber\\
\end{eqnarray}
\normalsize
where
\begin{eqnarray}
\label{15}
\textbf{B}^{\ast}=\textbf{B}+d^{2}_{e}\nabla\times\rho^{-1}\left(\nabla\times\textbf{B}\right).
\end{eqnarray}
For the simplicity, we consider a domain $S_3$ with periodic boundary conditions.  

The conservation of energy of the extended MHD was studied by \cite{K.Kimura}; the total energy is given as
\begin{eqnarray}
\label{16}
\mathscr{H}:=\int_{\Omega}\left\{\rho\left(\frac{V^{2}}{2}+U\left(\rho\right)\right)+\frac{B^{2}}{2}+d^{2}_{e}\frac{\left(\nabla\times\textbf{B}\right)^{2}}{2 \rho}\right\} d^{3}x.
\end{eqnarray}
This $\mathscr{H}$ is the natural candidate of the Hamiltonian.

To formulate the Hamiltonian system, we consider a \emph{phase space} spanned by the variables
$\rho, \textbf{V}$, and $\textbf{B}^{\ast}$; we denote the 
the state vector by $u=\left(\rho, \textbf{V},\textbf{B}^{\ast}\right)^{t}$. Then, $\textbf{B}$ in $\mathscr{H}$ must be evaluated as a function of $\textbf{B}^{\ast}$ and $\rho$ by (\ref{15}).
The gradient of the Hamiltonian $\mathscr{H}$ is 
\begin{eqnarray*}
\partial_u\mathscr{H}= \left(
\begin{array}{c}
	\partial_\rho\mathscr{H}\\
\\
	\partial_\bi{V}\mathscr{H}\\
	\\
\partial_{\bi{B}^\ast}\mathscr{H}	
\end{array}
\right)=\left(
\begin{array}{c}
	\frac{V^{2}}{2}+h+d^{2}_{e}\left(\frac{\left(\nabla\times\bi{B}\right)^{2}}{2\rho^{2}}\right)\\
	\\
\rho \bi{V}\\
	\\
\bi{B}	
\end{array}
\right).
\end{eqnarray*}

Now, we propose a Poisson operator for the extended MHD equations:
\small
\begin{eqnarray}
\label{17}
\fl\mathcal{J} = \left(
\begin{array}{ccc}
	0&-\nabla\cdot&0\\
	\\
	-\nabla&-\rho^{-1}\left(\nabla\times\textbf{V}\right)\times\circ&\rho^{-1}\left(\nabla\times\circ\right)\times\textbf{B}^{\ast}\\
	\\
0&\nabla\times\left(\circ\times\rho^{-1}\textbf{B}^{\ast}\right)&-d_{i}\nabla\times\left(\rho^{-1}\left(\nabla\times\circ\right)\times\textbf{B}^{\ast}\right)+d^{2}_{e}\nabla\times\left(\rho^{-1}\left(\nabla\times\circ\right)\times\left(\nabla\times\textbf{V}\right)\right)	
\end{array}
\right).\nonumber\\
\end{eqnarray}
\normalsize

With the Poisson operator (\ref{17}) and the Hamiltonian (\ref{16}), 
Hamilton's equation (\ref{11}) reproduces the extended MHD equations (\ref{12}), (\ref{13}), and (\ref{14}).

Using the periodic boundary conditions, we can easily demonstrate the antisymmetry of $\mathcal{J}$.
Hence the Poisson bracket defined by this $\mathcal{J}$ satisfies antisymmetry.
However, the proof of Jacobi's identity is rather elaborating.
Leaving it for the next section, we end this section by stating the main assertion:

\begin{theorem}[Poisson algebra of extended MHD]
\label{theorem}
We define a bilinear form
\[
\{ F, G \} = \langle \partial_{u} F, \mathcal{J} \partial_{u} G \rangle.
\]
Then, $\{ ~, ~ \}$ is a Poisson bracket, and $C^\infty_{\{~,~\}}(X)$ is a Poisson algebra.
Providing it with a Hamiltonian $\mathscr{H}$ of (16), we obtain the extended MHD system.
\end{theorem}

\section{Jacobi's Identity}
\subsection{Basic algebra}
\indent \indent 
We have yet to prove Jacobi's identity for the Poisson bracket.
Apparently, it is not of a Lie-Poisson type.
Complexity is cause by the factor $\rho^{-1}$, as well as differential operator $\nabla\times$ 
appearing in many places of $\mathcal{J}$.
However, there is a basic, common permutation relation that generates the total Poisson system.
We prepare the following lemma:

\begin{lemma}
\label{lemma:base_algebra}
On $C^\infty(X)$, we define a bracket (bilinear form) 
\begin{eqnarray}
\label{18}
\left[F,G\right]^\textbf{p}_{\textbf{q},\textbf{r}}=\int_{\Omega} \bigg[\rho^{-1}\left(\nabla\times \textbf{p}\right)\cdot \left(\partial_{\textbf{q}}F\times \partial_{\textbf{r}}G\right)-\partial_{\rho}F \left(\nabla\cdot \partial_{\textbf{r}}G\right)-\partial_{\textbf{q}}F\cdot\nabla \partial_{\rho}G\bigg]d^{3}x,
\end{eqnarray}
where $\textbf{p}, \textbf{q}$, and $\textbf{r}$ are vector fields arbitrarily chosen from $\textbf{V}$ or $\textbf{A}^*$ (where $\textbf{A}^*$is the vector potential and is related to $\textbf{B}^*$ by the relation $\textbf{B}^{\ast}=\nabla \times\textbf{A}^{\ast}$).
This bracket satisfies an antisymmetry relation
\[
\left[F,G\right]^{\textbf{p}}_{\textbf{q},\textbf{r}}=-\left[G,F\right]^{\textbf{p}}_{\textbf{r},\textbf{q}},
\]
as well as a permutation law 
\begin{eqnarray}\label{h1}
\left[E,\left[F,G\right]^{\textbf{p}}_{\textbf{q},\textbf{r}}\right]^{\textbf{p}}_{\textbf{s},\textbf{p}}+\left[G,\left[E,F\right]^{\textbf{p}}_{\textbf{s},\textbf{q}}\right]^{\textbf{p}}_{\textbf{r},\textbf{p}}+\left[F,\left[G,E\right]^{\textbf{p}}_{\textbf{r},\textbf{s}}\right]^{\textbf{p}}_{\textbf{q},\textbf{p}}= O(\partial^2),
\end{eqnarray}
where $O(\partial^2)$ denotes terms including second-order derivatives.
Hence, the sum over the permutation vanishes on the modulo operation by $\partial^2$.

\end{lemma}

The combination of the functionals $E,~F,~G$ and the corresponding state variables $\textbf{q},~\textbf{r},~\textbf{s}$ is a unique aspect of this bracket. Notice that the permutation law (\ref{h1}) resembles Jacobi's identity. In fact, the algebraic relation delineated by this Lemma \ref{lemma:base_algebra} is the root cause of Jacobi's identity satisfied by the Poisson bracket.\\

(proof of Lemma\,\ref{lemma:base_algebra})
The antisymmetry is evident.
To prove Jacobi's identity, we have to calculate the functional derivative of the bracket.
By the inhomogeneous factor $\rho^{-1}\left(\nabla\times \bi{p}\right)$ included in the bracket,
the derivatives such as $\partial_\rho  \left[F,G\right]^{\bi{p}}_{\bi{q},\bi{r}}$
and $\partial_{\bi{p}}  \left[F,G\right]^{\bi{p}}_{\bi{q},\bi{r}}$
are sums of the terms that consist of only first derivatives of $F$ and $G$, as well as the terms including
second-order derivatives 
(the second-order terms are modulo-outed in (\ref{h1})).
The former ones are such that
\begin{eqnarray*}
\partial_{\rho}\left[F,G\right]^{\textbf{p}}_{\textbf{q},\textbf{r}}=-\rho^{-2}\left(\nabla\times \textbf{p}\right)\cdot \left(\partial_{\textbf{q}}F\times \partial_{\textbf{r}}G\right)
+ O(\partial^2),
\end{eqnarray*}
\begin{eqnarray*}
\partial_{\textbf{p}}\left[F,G\right]^{\textbf{p}}_{\textbf{q},\textbf{r}}=\nabla\times \rho^{-1}\left(\partial_{\textbf{q}}F\times \partial_{\textbf{r}}G\right)
+ O(\partial^2).
\end{eqnarray*}
The permutation low is given as
\begin{eqnarray}
\label{h}
\fl\left[E,\left[F,G\right]^{\textbf{p}}_{\textbf{q},\textbf{r}}\right]^{\textbf{p}}_{\textbf{s},\textbf{p}}+\left[G,\left[E,F\right]^{\textbf{p}}_{\textbf{s},\textbf{q}}\right]^{\textbf{p}}_{\textbf{r},\textbf{p}}&+\left[F,\left[G,E\right]^{\textbf{p}}_{\textbf{r},\textbf{s}}\right]^{\textbf{p}}_{\textbf{q},\textbf{p}}=\nonumber \\ & \int_{\Omega}\bigg\{\left(\nabla\times \textbf{p}\right)\cdot \big[ \rho^{-1} \partial_{\textbf{s}}E\times\big[\nabla\times\big(\partial_{\textbf{q}}F\times \rho^{-1} \partial_{\textbf{r}}G\big)\big]\big]\nonumber \\ &  +\partial_{\textbf{s}}E\cdot\nabla\big[\left(\nabla\times\textbf{p}\right)\cdot\left( \rho^{-1} \partial_{\textbf{q}}F\times \rho^{-1} \partial_{\textbf{r}}G\right)\big]\nonumber \\ &+\left(\nabla\times \textbf{p}\right)\cdot \big[ \rho^{-1} \partial_{\textbf{r}}G\times\big[\nabla\times\big(\partial_{\textbf{s}}E\times \rho^{-1}\partial_{\textbf{q}}F\big)\big]\big]\nonumber \\ & + \partial_{\textbf{r}}G\cdot\nabla\big[\left(\nabla\times\textbf{p}\right)\cdot\left( \rho^{-1} \partial_{\textbf{s}}E\times \rho^{-1} \partial_{\textbf{q}}F\right)\big]\nonumber \\ &+\left(\nabla\times \textbf{p}\right)\cdot \big[ \rho^{-1} \partial_{\textbf{q}}F\times\big[\nabla\times\big(\partial_{\textbf{r}}G\times \rho^{-1}\partial_{\textbf{s}}E\big)\big]\big]\nonumber \\ &  + \partial_{\textbf{q}}F\cdot\nabla\big[\left(\nabla\times\textbf{p}\right)\cdot\left( \rho^{-1} \partial_{\textbf{r}}G\times \rho^{-1} \partial_{\textbf{s}}E\right)\big]\bigg\} d^{3}x+O(\partial^2).
\end{eqnarray}
Denoting $\textbf{e}:= \partial_{\textbf{s}}E$, etc., equation (\ref{h}) reads
\begin{eqnarray}
\fl\left[E,\left[F,G\right]^{\textbf{p}}_{\textbf{q},\textbf{r}}\right]^{\textbf{p}}_{\textbf{s},\textbf{p}}+\left[G,\left[E,F\right]^{\textbf{p}}_{\textbf{s},\textbf{q}}\right]^{\textbf{p}}_{\textbf{r},\textbf{p}}&+\left[F,\left[G,E\right]^{\textbf{p}}_{\textbf{r},\textbf{s}}\right]^{\textbf{p}}_{\textbf{q},\textbf{p}}=\nonumber \\ &\int_{\Omega} \big\{\left(\nabla\times \textbf{p}\right)\cdot \big[\rho^{-1} \textbf{e}\times\big[\nabla\times\big(\textbf{f}\times \rho^{-1} \textbf{g}\big)\big]\big]\nonumber \\ &+ \textbf{e}\cdot\nabla\big[\left(\nabla\times\textbf{p}\right)\cdot\left( \rho^{-1} \textbf{f}\times \rho^{-1}\textbf{g}\right)\big]+\circlearrowleft\big]\big\} d^{3}x+O(\partial^2)
\end{eqnarray}
where $\circlearrowleft$ denotes the summation over cyclic permutation of the vectors $\textbf{e}, \textbf{f},$ and $\textbf{g}$. After integrating by parts,
the integrand of $\left[E,\left[F,G\right]^{\textbf{p}}_{\textbf{q},\textbf{r}}\right]^{\textbf{p}}_{\textbf{s},\textbf{p}}$ can be written as
\begin{eqnarray}
\label{21}
\left(\nabla\times\textbf{p}\right)\cdot\big\{\rho^{-1}\textbf{e}\times\left[\nabla\times\left( \textbf{f}\times\rho^{-1}\textbf{g}\right)\right]-\left(\rho^{-1}\textbf{f}\times\rho^{-1}\textbf{g}\right)\nabla\cdot \textbf{e}\big\} .
\end{eqnarray}
The term bracketed by $\left\{~\right\}$ can be rewritten by vector identities as
\small
\begin{eqnarray*}
\rho^{-1}\textbf{e}\times \big[\textbf{f}\left(\nabla\cdot\rho^{-1}\textbf{g}\right)-\rho^{-1}\textbf{g}\left(\nabla\cdot \textbf{f}\right)+\left(\rho^{-1}\textbf{g}\cdot \nabla\right)\textbf{x}-\left(\textbf{f}\cdot \nabla\right)\rho^{-1} \textbf{g}\big]-\left(\rho^{-1}\textbf{f}\times\rho^{-1}\textbf{g}\right)\nabla\cdot \textbf{e}.
\end{eqnarray*}
\normalsize
The second term and the last term cancel by summation over permutations. 
To deal with the residual terms in (\ref{21}), we use the symmetry of the curl operator
\begin{eqnarray*}
\textbf{p}\cdot\nabla\times\bigg\{ \rho^{-1}\textbf{e}\times\big[\textbf{f}\left(\nabla\cdot\rho^{-1}\textbf{g}\right)+\left(\rho^{-1}\textbf{g}\cdot \nabla\right)\textbf{f}-\left(\textbf{f}\cdot \nabla\right)\rho^{-1} \textbf{g}\big]\bigg\} .
\end{eqnarray*}
Invoking Levi-Civita symbol, we may write
\begin{eqnarray}
\label{22}
&\epsilon_{ijk}\partial_{j}\bigg\{\epsilon_{klm}\rho^{-1}e_{l}\big[f_{m}\partial_{n}\left(\rho^{-1}g_{n}\right)+\rho^{-1}g_{n}\partial_{n}f_{m}-f_{n}\partial_{n}\left(\rho^{-1} g_{m}\right)\big]\bigg\}\nonumber\\\fl&=\partial_{j}\bigg\{\rho^{-1}e_{i}\big[\partial_{n}\left(\rho^{-1} g_{n}f_{j}\right)-f_{n}\partial_{n}\left(\rho^{-1}g_{j}\right)\big]-\rho^{-1}e_{j}\big[\partial_{n}\left(\rho^{-1}g_{n}f_{i}\right)-f_{n}\partial_{n}\left(\rho^{-1}g_{i}\right)\big]\bigg\} .
\end{eqnarray}
The last two terms are manipulated as
\begin{eqnarray*}
-\partial_{n}\left(\rho^{-2}g_{n}f_{i} e_{j}\right)+\rho^{-1}g_{n}f_{i}\partial_{n}\left(\rho^{-1}e_{j}\right)+\partial_{n}\left(\rho^{-2}g_{i}f_{n}e_{j}\right)-\rho^{-1}g_{i}\partial_{n}\left(\rho^{-1}f_{n}e_{j}\right) .
\end{eqnarray*}
Now (\ref{22}) is summarized as
\begin{eqnarray*}
\partial_{j}\partial_{n}\left[\rho^{-2}\left(g_{i}f_{n}e_{j}-g_{n}f_{i}e_{j}\right]\right)&+\partial_{j}\left[\rho^{-1}e_{i}\partial_{n}\left(\rho^{-1}g_{n}f_{j}\right)-\rho^{-1}g_{i}\partial_{n}\left(\rho^{-1}f_{n}e_{j}\right)\right]\nonumber\\&+\partial_{j}\left[\rho^{-1}g_{n}f_{i} \partial_{n}\left(\rho^{-1}e_{j}\right)-\rho^{-1}f_{n}e_{i} \partial_{n}\left(\rho^{-1}g_{j}\right)\right].\nonumber
\end{eqnarray*}
each term of which cancels out by summation over the permutation.
(QED)

\begin{remark}
If we choose $\textbf{p}=\textbf{q}=\textbf{r}=\textbf{V}$, 
the bracket (\ref{18}) is the Poisson bracket of the barotropic compressible fluid:
\begin{eqnarray}
\label{23}
\left\{F,G\right\}=\int_{\Omega} \bigg[\rho^{-1}\left(\nabla\times \textbf{V}\right)\cdot \left(\partial_\textbf{V}F\times \partial_{\textbf{V}}G\right)-\partial_{\rho}F \left(\nabla\cdot \partial_{\textbf{V}}G\right)-\partial_{\textbf{V}}F\cdot\nabla \partial_{\rho}G\bigg]d^{3}x ,
\end{eqnarray}
where the state vector is $u=\left(\rho,\textbf{V}\right)$.
The Poisson operator corresponding to Poisson bracket (\ref{23}) is
\begin{eqnarray}
\mathcal{J}= \left(
\begin{array}{cc}
	0&-\nabla\cdot\\
	\\
	-\nabla&-\rho^{-1}\left(\nabla\times\textbf{V}\right)\times\circ
\end{array}
\right).
\end{eqnarray}
Giving a Hamiltonian
\begin{eqnarray}
\mathscr{H}:=\int_{\Omega}\rho\left(\frac{V^{2}}{2}+U\left(\rho\right)\right) d^{3}x,
\end{eqnarray}
Hamilton's equation reads
\begin{eqnarray}
\frac{\partial \rho}{\partial t}=-\nabla\cdot\left(\rho\textbf{V}\right) ,
\end{eqnarray}
\begin{eqnarray}
\frac{\partial \textbf{V}}{\partial t}=-\left(\nabla\times\textbf{V}\right)\times\textbf{V}-\nabla\left(h+\frac{V^{2}}{2}\right).
\end{eqnarray}
\end{remark}

\subsection{Jacobi's identity for the Poisson bracket of extended MHD}
\indent \indent Now we complete the proof of Theorem \ref{theorem} by verifying Jacobi's identity for the Poisson bracket
\begin{eqnarray}
\label{28}
\left\{F,G\right\}= -\int_{\Omega}&\bigg\{\left[F_{\rho} \nabla\cdot G_{\textbf{V}}+F_{\textbf{V}}\cdot\nabla G_{\rho}\right]-\left[\rho^{-1}\left(\nabla\times\textbf{V}\right)\cdot\big(F_{\textbf{V}}\times G_{\textbf{V}}\big)\right]\nonumber \\ 
&-\left[\textbf{B}^{\ast}\cdot\rho^{-1}\big(F_{\textbf{V}}\times\left(\nabla\times G_{\textbf{B}^{\ast}}\right)\big)+\textbf{B}^{\ast}\cdot\rho^{-1}\big(\left(\nabla\times F_{\textbf{B}^{\ast}}\right)\times G_{\textbf{V}}\big)\right]\nonumber \\&+ d_{i} \left[ \textbf{B}^{\ast}\cdot\rho^{-1}\big(\left(\nabla\times F_{\textbf{B}^{\ast}}\right)\times\left(\nabla\times G_{\textbf{B}^{\ast}}\right)\big)\right]\nonumber \\ &
-d^{2}_{e}\left[ \left(\nabla\times\textbf{V}\right)\cdot\rho^{-1}\big(\left(\nabla\times F_{\textbf{B}^{\ast}}\right)\times\left(\nabla\times G_{\textbf{B}^{\ast}}\right)\big)\right]\bigg\}d^{3}x,
\end{eqnarray}
where the subscripts indicate functional derivative of the functional $F, G$ with respect to the state variables $\rho, \textbf{V}, \textbf{B}^{\ast}$, i.e $F_{\rho}= \partial_{\rho} F$. 


To examine Jacobi's identity, we have to study the derivatives of the bracket by the state variables,
which consists of two groups of terms;
group (A) is the collection of terms that include second-order derivatives (such as $F_{\textbf{B}^{\ast},\textbf{V}}$).
Formally, group (A) is generated by pretending that the coefficients in the Poisson operator $\mathcal{J}$
are independent to (or, different from) the state vector $u$. The terms of group (A) cancel out when summed up in 
$\left\{E,\left\{F,G\right\}\right\}+\circlearrowleft$ \footnote{Suppose that a bracket is defined by
\begin{eqnarray*}
\left\{F,G\right\}=\left\langle \partial_{u}F, \mathcal{J} \partial_{u}G\right\rangle
\end{eqnarray*}
with an antisymetric ($\left\langle \textbf{u},\mathcal{J}\textbf{v}\right\rangle=-\left\langle \textbf{v},\mathcal{J}\textbf{u}\right\rangle$), constant ($\partial_{u}\mathcal{J}$=0) operator $\mathcal{J}$. Then $\left\{F,G\right\}$ is a Poisson bracket(i.e.  antisymmetry, Jacobi’s identity, and  Leibniz property hold); cf.\cite{Morrison82}.}.
Group (B) summarizes the remaining terms that are due to the derivatives of $\mathcal{J}$ by $u$;
explicitly, we have

\begin{eqnarray}
\label{30}
\left\{F,G\right\}_{\rho} ~\mathrm{mod}~\partial^2
=&-\rho^{-2}\left(\nabla\times\textbf{V}\right)\cdot\left(F_{\textbf{V}}\times G_{\textbf{V}}\right)\nonumber\\&- \rho^{-2}\textbf{B}^{\ast}\cdot \left[F_{\textbf{V}}\times\left(\nabla\times G_{\textbf{B}^{\ast}}\right)\right]\nonumber\\&-\rho^{-2}\textbf{B}^{\ast}\cdot \left[\left(\nabla\times F_{\textbf{B}^{\ast}}\right)\times G_{\textbf{V}}\right]\nonumber\\&+d_{i} \bigg[\rho^{-2}\textbf{B}^{\ast}\cdot \left[\left(\nabla\times F_{\textbf{B}^{\ast}}\right)\times\left(\nabla\times G_{\textbf{B}^{\ast}}\right)\right]\bigg]\nonumber\\&-d^{2}_{e} \bigg[\rho^{-2}\left(\nabla\times\textbf{V}\right)\cdot \left[\left(\nabla\times F_{\textbf{B}^{\ast}}\right)\times\left(\nabla\times G_{\textbf{B}^{\ast}}\right)\right]\bigg],
\end{eqnarray}

\begin{eqnarray}
\label{31}
\left\{F,G\right\}_{\textbf{V}}  ~\mathrm{mod}~\partial^2
=&\nabla\times\rho^{-1}\left(F_{\textbf{V}}\times G_{\textbf{V}}\right)\nonumber\\&+d^{2}_{e} \bigg[\nabla\times\rho^{-1}\big(\left(\nabla\times F_{\textbf{B}^{\ast}}\right)\times\left(\nabla\times G_{\textbf{B}^{\ast}}\right)\big)\bigg],
\end{eqnarray}

\begin{eqnarray}
\label{32}
\left\{F,G\right\}_{\textbf{B}^{\ast}}  ~\mathrm{mod}~\partial^2
= &\rho^{-1}\big(F_{\textbf{V}}\times \left(\nabla\times G_{\textbf{B}^{\ast}}\right)\big)\nonumber\\&+\rho^{-1}\big(\left(\nabla\times F_{\textbf{B}^{\ast}}\right)\times G_{\textbf{V}}\big)\nonumber\\&-d_{i} \bigg[\rho^{-1}\big(\left(\nabla\times F_{\textbf{B}^{\ast}}\right)\times\left(\nabla\times G_{\textbf{B}^{\ast}}\right)\big)\bigg].
\end{eqnarray}
In what follows, we show that the remaining group (B) terms cancel out.
By (\ref{30}), (\ref{31}), and (\ref{32}), we obtain
\small
\begin{eqnarray}
\fl\left\{E,\left\{F,G\right\}\right\}+\circlearrowleft
=&\int_{\Omega} E_{\textbf{V}}\cdot \bigg[\nabla \bigg(\rho^{-2}\left(\nabla\times\textbf{V}\right)\cdot\big(F_{\textbf{V}}\times G_{\textbf{V}}\big)\bigg)\nonumber\\&- \rho^{-1}\left(\nabla\times\textbf{V}\right)\times\big[\nabla\times\rho^{-1}\left(F_{\textbf{V}}\times G_{\textbf{V}}\right)\big]\bigg]d^{3}x\nonumber\\&
+\int_{\Omega} E_{\textbf{V}}\cdot \bigg[\nabla \bigg(\rho^{-2}\textbf{B}^{\ast}\cdot \left[F_{\textbf{V}}\times\left(\nabla\times G_{\textbf{B}^{\ast}}\right)\right]\bigg)\nonumber\\&+ \big[\nabla\times\big(\rho^{-1}F_{\textbf{V}}\times \left(\nabla\times G_{\textbf{B}^{\ast}}\right)\big)\big]\times \rho^{-1}\textbf{B}^{\ast}\bigg]d^{3}x\nonumber\\&
+\int_{\Omega} E_{\textbf{V}}\cdot \bigg[\nabla \bigg(\rho^{-2}\textbf{B}^{\ast}\cdot \left[\left(\nabla\times F_{\textbf{B}^{\ast}}\right)\times G_{\textbf{V}}\right]\bigg)\nonumber\\&+ \big[\nabla\times\big(\rho^{-1}\left(\nabla\times F_{\textbf{B}^{\ast}}\right)\times G_{\textbf{V}}\big)\big]\times \rho^{-1}\textbf{B}^{\ast} \bigg]d^{3}x\nonumber\\&
-d_{i}\int_{\Omega} E_{\textbf{V}}\cdot \bigg[\nabla \bigg(\rho^{-2}\textbf{B}^{\ast}\cdot \left[\left(\nabla\times F_{\textbf{B}^{\ast}}\right)\times\left(\nabla\times G_{\textbf{B}^{\ast}}\right)\right]\bigg)\nonumber\\&+\big[\nabla\times\big(\rho^{-1}\left(\nabla\times F_{\textbf{B}^{\ast}}\right)\times\left(\nabla\times G_{\textbf{B}^{\ast}}\right)\big)\big]\times\rho^{-1}\textbf{B}^{\ast}\bigg]d^{3}x\nonumber\\&
+d^{2}_{e}\int_{\Omega} E_{\textbf{V}}\cdot \bigg[\nabla \bigg(\rho^{-2}\left(\nabla\times\textbf{V}\right)\cdot \left[\left(\nabla\times F_{\textbf{B}^{\ast}}\right)\times\left(\nabla\times G_{\textbf{B}^{\ast}}\right)\right]\bigg)\nonumber\\&-\rho^{-1}\left(\nabla\times\textbf{V}\right)\times \big[\nabla\times\big(\rho^{-1}\left(\nabla\times F_{\textbf{B}^{\ast}}\right)\times\left(\nabla\times G_{\textbf{B}^{\ast}}\right)\big)\big]\bigg]d^{3}x\nonumber\\&
-d_{i} \int_{\Omega} E_{\textbf{B}^{\ast}}\cdot \nabla\times\bigg[\big[\nabla\times\big(\rho^{-1}F_{\textbf{V}}\times\left(\nabla\times G_{\textbf{B}^{\ast}}\right)\big)\big]\times \rho^{-1}\textbf{B}^{\ast} \nonumber\\&+ \big[\nabla\times\big(\rho^{-1}\left(\nabla\times F_{\textbf{B}^{\ast}}\right)\times G_{\textbf{V}}\big)\big]\times \rho^{-1}\textbf{B}^{\ast}\bigg] d^{3}x\nonumber\\&
+d^{2}_{e} \int_{\Omega} E_{\textbf{B}^{\ast}}\cdot \nabla\times\bigg[\big[\nabla\times\big(\rho^{-1}F_{\textbf{V}}\times\left(\nabla\times G_{\textbf{B}^{\ast}}\right)\big)\big]\times \rho^{-1} \left(\nabla\times\textbf{V}\right)\nonumber\\&+ \big[\nabla\times\big(\rho^{-1}\left(\nabla\times F_{\textbf{B}^{\ast}}\right)\times G_{\textbf{V}}\big)\big]\times \rho^{-1}\left(\nabla\times\textbf{V}\right)\bigg] d^{3}x\nonumber\\&
+\int_{\Omega} E_{\textbf{B}^{\ast}}\cdot \nabla\times\bigg[\big[\nabla\times\rho^{-1}\left(F_{\textbf{V}}\times G_{\textbf{V}}\right)\big]\times\rho^{-1}\textbf{B}^{\ast}\bigg]d^{3}x\nonumber\\&
+d^{2}_{e}\int_{\Omega} E_{\textbf{B}^{\ast}}\cdot \nabla\times\bigg[\big[\nabla\times\big(\rho^{-1}\left(\nabla\times F_{\textbf{B}^{\ast}}\right)\times\left(\nabla\times G_{\textbf{B}^{\ast}}\right)\big)\big]\times\rho^{-1}\textbf{B}^{\ast}\bigg]d^{3}x\nonumber\\&
+d^{2}_{i} \int_{\Omega} E_{\textbf{B}^{\ast}}\cdot\nabla\times\bigg[\big[\nabla\times\big(\rho^{-1}\left(\nabla\times F_{\textbf{B}^{\ast}}\right)\times\left(\nabla\times G_{\textbf{B}^{\ast}}\right)\big)\big]\times\rho^{-1}\textbf{B}^{\ast}\bigg]d^{3}x\nonumber\\&
-d_{i}d^{2}_{e}\int_{\Omega} E_{\textbf{B}^{\ast}}\cdot\nabla\times\bigg[\big[\nabla\times\rho^{-1}\big(\left(\nabla\times F_{\textbf{B}^{\ast}}\right)\times\left(\nabla\times G_{\textbf{B}^{\ast}}\right)\big)\big]\times\rho^{-1}\left(\nabla\times\textbf{V}\right)\bigg]d^{3}x+\circlearrowleft  
\nonumber\\& ~~~+ O(\partial^2).
\nonumber\\
\end{eqnarray}
\normalsize
To prove Jacobi's identity, we collect terms that have the same combinations of functional derivatives such that $\left(E_{\textbf{V}},F_{\textbf{V}},G_{\textbf{V}}\right)$, $\left(E_{\textbf{V}},F_{\textbf{V}},G_{\textbf{B}^{\ast}}\right)$$,...,$$\left(E_{\textbf{B}^{\ast}},F_{\textbf{B}^{\ast}},G_{\textbf{B}^{\ast}}\right)$. 
Then, 
\small
\begin{eqnarray}
\label{35}
\fl \left\{E,\left\{F,G\right\}\right\}+\circlearrowleft&
=\int_{\Omega} \big\{\left(\nabla\times\textbf{V}\right)\cdot\big[\rho^{-1}E_{\textbf{V}}\times\big(\nabla\times\rho^{-1}\left(F_{\textbf{V}}\times G_{\textbf{V}}\right)\big)\big]+E_{\textbf{V}}\cdot \big[\nabla \big(\rho^{-2}\left(\nabla\times\textbf{V}\right)\cdot\left(F_{\textbf{V}}\times G_{\textbf{V}}\right)\big)\big]\nonumber\\&  
+\left(\nabla\times\textbf{V}\right)\cdot\big[\rho^{-1}G_{\textbf{V}}\times\big(\nabla\times\rho^{-1}\left(E_{\textbf{V}}\times F_{\textbf{V}}\right)\big)\big]+G_{\textbf{V}}\cdot \big[\nabla \big(\rho^{-2}\left(\nabla\times\textbf{V}\right)\cdot\left(E_{\textbf{V}}\times F_{\textbf{V}}\right)\big)\big]\nonumber\\&
+ \left(\nabla\times\textbf{V}\right)\cdot\big[\rho^{-1}F_{\textbf{V}}\times\big(\nabla\times\rho^{-1}\left(G_{\textbf{V}}\times E_{\textbf{V}}\right)\big)\big]+F_{\textbf{V}}\cdot \big[\nabla \big(\rho^{-2}\left(\nabla\times\textbf{V}\right)\cdot\left(G_{\textbf{V}}\times E_{\textbf{V}}\right)\big)\big]\big\}d^{3}x\nonumber\\&
+\int_{\Omega} \big\{\textbf{B}^{\ast}\cdot\big[\rho^{-1}\left(\nabla\times E_{\textbf{B}^{\ast}}\right)\times\big(\nabla\times\rho^{-1}\left[  F_{\textbf{V}}\times G_{\textbf{V}}\right]\big)\big]\nonumber\\&  
+ \textbf{B}^{\ast}\cdot\big[\rho^{-1}G_{\textbf{V}}\times\big(\nabla\times\rho^{-1}\left[\left(\nabla\times E_{\textbf{B}^{\ast}}\right)\times F_{\textbf{V}}\right]\big)\big]+G_{\textbf{V}}\cdot \nabla \big[\rho^{-2}\textbf{B}^{\ast}\cdot \big( \left(\nabla\times E_{\textbf{B}^{\ast}}\right)\times F_{\textbf{V}}\big)\big]\nonumber\\& 
+\textbf{B}^{\ast}\cdot\big[\rho^{-1}F_{\textbf{V}}\times\big(\nabla\times\rho^{-1}\left[G_{\textbf{V}}\times \left(\nabla\times E_{\textbf{B}^{\ast}}\right)\right]\big)\big]+F_{\textbf{V}}\cdot \nabla \big[\rho^{-2}\textbf{B}^{\ast}\cdot \big( G_{\textbf{V}}\times\left(\nabla\times E_{\textbf{B}^{\ast}}\right)\big)\big]
\big\}d^{3}x\nonumber\\&
+\int_{\Omega} \big\{\textbf{B}^{\ast}\cdot\big[\rho^{-1}E_{\textbf{V}}\times\big(\nabla\times\rho^{-1}\left[F_{\textbf{V}}\times \left(\nabla\times G_{\textbf{B}^{\ast}}\right)\right]\big)\big]+E_{\textbf{V}}\cdot \nabla \big[\rho^{-2}\textbf{B}^{\ast}\cdot \big(F_{\textbf{V}}\times \left(\nabla\times G_{\textbf{B}^{\ast}}\right)\big)\big]\nonumber\\&  
+\textbf{B}^{\ast}\cdot\big[\rho^{-1}\left(\nabla\times G_{\textbf{B}^{\ast}}\right)\times\big(\nabla\times\rho^{-1}\left[E_{\textbf{V}}\times G_{\textbf{V}}\right]\big)\big]\nonumber\\&
+ \textbf{B}^{\ast}\cdot\big[\rho^{-1}F_{\textbf{V}}\times\big(\nabla\times\rho^{-1}\left[\left(\nabla\times G_{\textbf{B}^{\ast}}\right)\times E_{\textbf{V}}\right]\big)\big]+F_{\textbf{V}}\cdot \nabla \big[\rho^{-2}\textbf{B}^{\ast}\cdot \big(\left(\nabla\times G_{\textbf{B}^{\ast}}\right)\times E_{\textbf{V}}\big)\big] \big\}d^{3}x
\nonumber\\&
+\int_{\Omega} \big\{\textbf{B}^{\ast}\cdot\big[\rho^{-1}E_{\textbf{V}}\times\big(\nabla\times\rho^{-1}\left[ \left(\nabla\times F_{\textbf{B}^{\ast}}\right)\times G_{\textbf{V}}\right]\big)\big]+E_{\textbf{V}}\cdot \nabla \big[\rho^{-2}\textbf{B}^{\ast}\cdot \big(\left(\nabla\times F_{\textbf{B}^{\ast}}\right)\times G_{\textbf{V}}\big)\big]\nonumber\\&  
+ \textbf{B}^{\ast}\cdot\big[\rho^{-1}G_{\textbf{V}}\times\big(\nabla\times\rho^{-1}\left[ E_{\textbf{V}}\times\left(\nabla\times F_{\textbf{B}^{\ast}}\right)\right]\big)\big]+G_{\textbf{V}}\cdot \nabla \big[\rho^{-2}\textbf{B}^{\ast}\cdot \big( E_{\textbf{V}}\times \left(\nabla\times F_{\textbf{B}^{\ast}}\right)\big)\big]\nonumber\\& 
+\textbf{B}^{\ast}\cdot\big[\rho^{-1}\left(\nabla\times F_{\textbf{B}^{\ast}}\right)\times\big(\nabla\times\rho^{-1}\left[G_{\textbf{V}}\times E_{\textbf{V}}\right]\big)\big]
\big\}d^{3}x\nonumber\\&
-d_{i}\int_{\Omega} \big\{\textbf{B}^{\ast}\cdot\big[\rho^{-1}E_{\textbf{V}}\times\big(\nabla\times\rho^{-1}\left[ \left(\nabla\times F_{\textbf{B}^{\ast}}\right)\times \left(\nabla\times G_{\textbf{B}^{\ast}}\right)\right]\big)\big]\nonumber\\&+E_{\textbf{V}}\cdot \nabla \big[\rho^{-2}\textbf{B}^{\ast}\cdot \big(\left(\nabla\times F_{\textbf{B}^{\ast}}\right)\times \left(\nabla\times G_{\textbf{B}^{\ast}}\right)\big)\big]\nonumber\\&  
+\textbf{B}^{\ast}\cdot \big[\rho^{-1}\left(\nabla\times G_{\textbf{B}^{\ast}}\right)\times \big(\nabla\times \rho^{-1}\left[E_{\textbf{V}}\times\left(\nabla\times F_{\textbf{B}^{\ast}}\right) \right]\big)\big]\nonumber\\&+\textbf{B}^{\ast}\cdot \big[\rho^{-1}\left(\nabla\times F_{\textbf{B}^{\ast}}\right)\times \big(\nabla\times \rho^{-1}\left[\left(\nabla\times F_{\textbf{B}^{\ast}}\right)\times E_{\textbf{V}} \right]\big)\big]\big\}d^{3}x\nonumber\\&
+d^{2}_{e}\int_{\Omega} \big\{\left(\nabla\times \textbf{V}\right)\cdot\big[\rho^{-1}E_{\textbf{V}}\times\big(\nabla\times\rho^{-1}\left[ \left(\nabla\times F_{\textbf{B}^{\ast}}\right)\times \left(\nabla\times G_{\textbf{B}^{\ast}}\right)\right]\big)\big]\nonumber\\&+E_{\textbf{V}}\cdot \nabla \big[\rho^{-2}\left(\nabla\times \textbf{V}\right)\cdot \big(\left(\nabla\times F_{\textbf{B}^{\ast}}\right)\times \left(\nabla\times G_{\textbf{B}^{\ast}}\right)\big)\big]\nonumber\\&  
+\left(\nabla\times \textbf{V}\right)\cdot \big[\rho^{-1}\left(\nabla\times G_{\textbf{B}^{\ast}}\right)\times \big(\nabla\times \rho^{-1}\left[E_{\textbf{V}}\times\left(\nabla\times F_{\textbf{B}^{\ast}}\right) \right]\big)\big]\nonumber\\&+\left(\nabla\times \textbf{V}\right)\cdot \big[\rho^{-1}\left(\nabla\times F_{\textbf{B}^{\ast}}\right)\times \big(\nabla\times \rho^{-1}\left[\left(\nabla\times F_{\textbf{B}^{\ast}}\right)\times E_{\textbf{V}} \right]\big)\big]\big\}d^{3}x\nonumber\\&
-d_{i}\int_{\Omega} \big\{\textbf{B}^{\ast}\cdot\big[\rho^{-1}\left(\nabla\times E_{\textbf{B}^{\ast}}\right)\times\big(\nabla\times\rho^{-1}\left[F_{\textbf{V}}\times \left(\nabla\times G_{\textbf{B}^{\ast}}\right)\right]\big)\big]\nonumber\\& 
+\textbf{B}^{\ast}\cdot \big[\rho^{-1}\left(\nabla\times G_{\textbf{B}^{\ast}}\right)\times \big(\nabla\times \rho^{-1}\left[\left(\nabla\times E_{\textbf{B}^{\ast}}\right)\times F_{\textbf{V}} \right]\big)\big]\nonumber\\& 
+\textbf{B}^{\ast}\cdot \big[\rho^{-1}F_{\textbf{V}}\times \big(\nabla\times \rho^{-1}\left[\left(\nabla\times G_{\textbf{B}^{\ast}}\right)\times \left(\nabla\times E_{\textbf{B}^{\ast}}\right) \right]\big)\big]\nonumber\\&+ F_{\textbf{V}}\cdot \nabla \big[\rho^{-2}\textbf{B}^{\ast}\cdot \big(\left(\nabla\times G_{\textbf{B}^{\ast}}\right)\times\left(\nabla\times E_{\textbf{B}^{\ast}}\right)\big)\big]\big\}d^{3}x\nonumber\\&
+d^{2}_{e}\int_{\Omega} \big\{\left(\nabla\times \textbf{V}\right)\cdot\big[\rho^{-1}\left(\nabla\times E_{\textbf{B}^{\ast}}\right)\times\big(\nabla\times\rho^{-1}\left[F_{\textbf{V}}\times \left(\nabla\times G_{\textbf{B}^{\ast}}\right)\right]\big)\big] \nonumber\\&
+\left(\nabla\times \textbf{V}\right)\cdot \big[\rho^{-1}\left(\nabla\times G_{\textbf{B}^{\ast}}\right)\times \big(\nabla\times \rho^{-1}\left[\left(\nabla\times E_{\textbf{B}^{\ast}}\right)\times F_{\textbf{V}} \right]\big)\big]\nonumber\\& 
+\left(\nabla\times \textbf{V}\right)\cdot \big[\rho^{-1}F_{\textbf{V}}\times \big(\nabla\times \rho^{-1}\left[\left(\nabla\times G_{\textbf{B}^{\ast}}\right)\times \left(\nabla\times E_{\textbf{B}^{\ast}}\right) \right]\big)\big]\nonumber\\&+ F_{\textbf{V}}\cdot \nabla \big[\rho^{-2}\left(\nabla\times\textbf{V}\right)\cdot \big(\left(\nabla\times G_{\textbf{B}^{\ast}}\right)\times\left(\nabla\times E_{\textbf{B}^{\ast}}\right)\big)\big]\bigg\}d^{3}x\nonumber\\&
-d_{i}\int_{\Omega} \bigg\{\textbf{B}^{\ast}\cdot\big[\rho^{-1}\left(\nabla\times E_{\textbf{B}^{\ast}}\right)\times\big(\nabla\times\rho^{-1}\left[\left(\nabla\times F_{\textbf{B}^{\ast}}\right)\times G_{\textbf{V}}\right]\big)\big] \nonumber\\&
+\textbf{B}^{\ast}\cdot \big[\rho^{-1}G_{\textbf{V}}\times \big(\nabla\times \rho^{-1}\left[\left(\nabla\times E_{\textbf{B}^{\ast}}\right)\times \left(\nabla\times F_{\textbf{B}^{\ast}}\right) \right]\big)\big]\nonumber\\& +G_{\textbf{V}}\cdot \nabla \big[\rho^{-2}\textbf{B}^{\ast}\cdot \big(\left(\nabla\times E_{\textbf{B}^{\ast}}\right)\times \left(\nabla\times F_{\textbf{B}^{\ast}}\right)\big)\big]\nonumber\\&
+\textbf{B}^{\ast}\cdot \big[\rho^{-1}\left(\nabla\times F_{\textbf{B}^{\ast}}\right)\times \big(\nabla\times \rho^{-1}\left[G_{\textbf{V}}\times \left(\nabla\times E_{\textbf{B}^{\ast}}\right) \right]\big)\big]\big\}d^{3}x\nonumber\\&
+d^{2}_{e}\int_{\Omega} \big\{\left(\nabla\times \textbf{V}\right)\cdot\big[\rho^{-1}\left(\nabla\times E_{\textbf{B}^{\ast}}\right)\times\big(\nabla\times\rho^{-1}\left[\left(\nabla\times F_{\textbf{B}^{\ast}}\right)\times G_{\textbf{V}}\right]\big)\big] \nonumber\\&
+\left(\nabla\times \textbf{V}\right)\cdot \big[\rho^{-1}G_{\textbf{V}}\times \big(\nabla\times \rho^{-1}\left[\left(\nabla\times E_{\textbf{B}^{\ast}}\right)\times \left(\nabla\times F_{\textbf{B}^{\ast}}\right) \right]\big)\big]\nonumber\\& +G_{\textbf{V}}\cdot \nabla \big[\rho^{-2}\left(\nabla\times \textbf{V}\right)\cdot \big(\left(\nabla\times E_{\textbf{B}^{\ast}}\right)\times \left(\nabla\times F_{\textbf{B}^{\ast}}\right)\big)\big]\nonumber\\&
+\left(\nabla\times \textbf{V}\right)\cdot \big[\rho^{-1}\left(\nabla\times F_{\textbf{B}^{\ast}}\right)\times \big(\nabla\times \rho^{-1}\left[G_{\textbf{V}}\times \left(\nabla\times E_{\textbf{B}^{\ast}}\right) \right]\big)\big]\big\}d^{3}x\nonumber\\&
+d^{2}_{i}\int_{\Omega} \big\{\textbf{B}^{\ast}\cdot\big[\rho^{-1}\left(\nabla\times E_{\textbf{B}^{\ast}}\right)\times\big(\nabla\times\rho^{-1}\left[\left(\nabla\times F_{\textbf{B}^{\ast}}\right)\times \left(\nabla\times G_{\textbf{B}^{\ast}}\right)\right]\big)\big]\nonumber\\&+\textbf{B}^{\ast}\cdot\big[\rho^{-1}\left(\nabla\times G_{\textbf{B}^{\ast}}\right)\times\big(\nabla\times\rho^{-1}\left[\left(\nabla\times E_{\textbf{B}^{\ast}}\right)\times \left(\nabla\times F_{\textbf{B}^{\ast}}\right)\right]\big)\big]\nonumber\\&+\textbf{B}^{\ast}\cdot\big[\rho^{-1}\left(\nabla\times F_{\textbf{B}^{\ast}}\right)\times\big(\nabla\times\rho^{-1}\left[\left(\nabla\times G_{\textbf{B}^{\ast}}\right)\times \left(\nabla\times E_{\textbf{B}^{\ast}}\right)\right]\big)\big]\big\}d^{3}x\nonumber\\&
+d^{2}_{e}\int_{\Omega} \big\{\textbf{B}^{\ast}\cdot\big[\rho^{-1}\left(\nabla\times E_{\textbf{B}^{\ast}}\right)\times\big(\nabla\times\rho^{-1}\left[\left(\nabla\times F_{\textbf{B}^{\ast}}\right)\times \left(\nabla\times G_{\textbf{B}^{\ast}}\right)\right]\big)\big]\nonumber\\&+\textbf{B}^{\ast}\cdot\big[\rho^{-1}\left(\nabla\times G_{\textbf{B}^{\ast}}\right)\times\big(\nabla\times\rho^{-1}\left[\left(\nabla\times E_{\textbf{B}^{\ast}}\right)\times \left(\nabla\times F_{\textbf{B}^{\ast}}\right)\right]\big)\big]\nonumber\\&+\textbf{B}^{\ast}\cdot\big[\rho^{-1}\left(\nabla\times F_{\textbf{B}^{\ast}}\right)\times\big(\nabla\times\rho^{-1}\left[\left(\nabla\times G_{\textbf{B}^{\ast}}\right)\times \left(\nabla\times E_{\textbf{B}^{\ast}}\right)\right]\big)\big]\big\}d^{3}x\nonumber\\&
-d_{i}d^{2}_{e}\int_{\Omega} \big\{\left(\nabla\times\textbf{V}\right)\cdot\big[\rho^{-1}\left(\nabla\times E_{\textbf{B}^{\ast}}\right)\times\big(\nabla\times\rho^{-1}\left[\left(\nabla\times F_{\textbf{B}^{\ast}}\right)\times \left(\nabla\times G_{\textbf{B}^{\ast}}\right)\right]\big)\big]\nonumber\\&+\left(\nabla\times\textbf{V}\right)\cdot\big[\rho^{-1}\left(\nabla\times G_{\textbf{B}^{\ast}}\right)\times\big(\nabla\times\rho^{-1}\left[\left(\nabla\times E_{\textbf{B}^{\ast}}\right)\times \left(\nabla\times F_{\textbf{B}^{\ast}}\right)\right]\big)\big]\nonumber\\&+\left(\nabla\times\textbf{V}\right)\cdot\big[\rho^{-1}\left(\nabla\times F_{\textbf{B}^{\ast}}\right)\times\big(\nabla\times\rho^{-1}\left[\left(\nabla\times G_{\textbf{B}^{\ast}}\right)\times \left(\nabla\times E_{\textbf{B}^{\ast}}\right)\right]\big)\big]\big\}d^{3}x
\nonumber\\& ~~~+ O(\partial^2).
\nonumber\\
\end{eqnarray}
\normalsize
To apply lemma\ref{lemma:base_algebra}, we rewrite (\ref{35}) in terms of the bilinear form (\ref{18}):
\small
\begin{eqnarray}
\fl\left\{E,\left\{F,G\right\}\right\}+\circlearrowleft&=\left[E,\left[F,G\right]^{\textbf{V}}_{\textbf{V},\textbf{V}}\right]^{\textbf{V}}_{\textbf{V},\textbf{V}}+\left[G,\left[E,F\right]^{\textbf{V}}_{\textbf{V},\textbf{V}}\right]^{\textbf{V}}_{\textbf{V},\textbf{V}}+\left[F,\left[G,E\right]^{\textbf{V}}_{\textbf{V},\textbf{V}}\right]^{\textbf{V}}_{\textbf{V},\textbf{V}}\nonumber\\&
+\left[E,\left[F,G\right]^{\textbf{A}^{\ast}}_{\textbf{V},\textbf{V}}\right]^{\textbf{A}^{\ast}}_{\textbf{A}^{\ast},\textbf{A}^{\ast}}+\left[G,\left[E,F\right]^{\textbf{A}^{\ast}}_{\textbf{A}^{\ast},\textbf{V}}\right]^{\textbf{A}^{\ast}}_{\textbf{V},\textbf{A}^{\ast}}+\left[F,\left[G,E\right]^{\textbf{A}^{\ast}}_{\textbf{V},\textbf{A}^{\ast}}\right]^{\textbf{A}^{\ast}}_{\textbf{V},\textbf{A}^{\ast}}\nonumber\\&
+\left[E,\left[F,G\right]^{\textbf{A}^{\ast}}_{\textbf{V},\textbf{A}^{\ast}}\right]^{\textbf{A}^{\ast}}_{\textbf{V},\textbf{A}^{\ast}}+\left[G,\left[E,F\right]^{\textbf{A}^{\ast}}_{\textbf{V},\textbf{V}}\right]^{\textbf{A}^{\ast}}_{\textbf{A}^{\ast},\textbf{A}^{\ast}}+\left[F,\left[G,E\right]^{\textbf{A}^{\ast}}_{\textbf{A}^{\ast},\textbf{V}}\right]^{\textbf{A}^{\ast}}_{\textbf{V},\textbf{A}^{\ast}}\nonumber\\&
+\left[E,\left[F,G\right]^{\textbf{A}^{\ast}}_{\textbf{A}^{\ast},\textbf{V}}\right]^{\textbf{A}^{\ast}}_{\textbf{V},\textbf{A}^{\ast}}+\left[G,\left[E,F\right]^{\textbf{A}^{\ast}}_{\textbf{V},\textbf{A}^{\ast}}\right]^{\textbf{A}^{\ast}}_{\textbf{V},\textbf{A}^{\ast}}+\left[F,\left[G,E\right]^{\textbf{A}^{\ast}}_{\textbf{V},\textbf{V}}\right]^{\textbf{A}^{\ast}}_{\textbf{A}^{\ast},\textbf{A}^{\ast}}\nonumber\\&
-d_{i}\bigg[\left[E,\left[F,G\right]^{\textbf{A}^{\ast}}_{\textbf{A}^{\ast},\textbf{A}^{\ast}}\right]^{\textbf{A}^{\ast}}_{\textbf{V},\textbf{A}^{\ast}}+\left[G,\left[E,F\right]^{\textbf{A}^{\ast}}_{\textbf{V},\textbf{A}^{\ast}}\right]^{\textbf{A}^{\ast}}_{\textbf{A}^{\ast},\textbf{A}^{\ast}}+\left[F,\left[G,E\right]^{\textbf{A}^{\ast}}_{\textbf{A}^{\ast},\textbf{V}}\right]^{\textbf{A}^{\ast}}_{\textbf{A}^{\ast},\textbf{A}^{\ast}}\bigg]\nonumber\\&
+d^{2}_{e}\bigg[\left[E,\left[F,G\right]^{\textbf{V}}_{\textbf{A}^{\ast},\textbf{A}^{\ast}}\right]^{\textbf{V}}_{\textbf{V},\textbf{V}}+\left[G,\left[E,F\right]^{\textbf{V}}_{\textbf{V},\textbf{A}^{\ast}}\right]^{\textbf{V}}_{\textbf{A}^{\ast},\textbf{V}}+\left[F,\left[G,E\right]^{\textbf{V}}_{\textbf{A}^{\ast},\textbf{V}}\right]^{\textbf{V}}_{\textbf{A}^{\ast},\textbf{V}}\bigg]\nonumber\\&
-d_{i}\bigg[\left[E,\left[F,G\right]^{\textbf{A}^{\ast}}_{\textbf{A}^{\ast},\textbf{V}}\right]^{\textbf{A}^{\ast}}_{\textbf{A}^{\ast},\textbf{A}^{\ast}}+\left[G,\left[E,F\right]^{\textbf{A}^{\ast}}_{\textbf{A}^{\ast},\textbf{A}^{\ast}}\right]^{\textbf{A}^{\ast}}_{\textbf{V},\textbf{A}^{\ast}}+\left[F,\left[G,E\right]^{\textbf{A}^{\ast}}_{\textbf{V},\textbf{A}^{\ast}}\right]^{\textbf{A}^{\ast}}_{\textbf{A}^{\ast},\textbf{A}^{\ast}}\bigg]\nonumber\\&
+d^{2}_{e}\bigg[\left[E,\left[F,G\right]^{\textbf{V}}_{\textbf{A}^{\ast},\textbf{V}}\right]^{\textbf{V}}_{\textbf{A}^{\ast},\textbf{V}}+\left[G,\left[E,F\right]^{\textbf{V}}_{\textbf{A}^{\ast},\textbf{A}^{\ast}}\right]^{\textbf{V}}_{\textbf{V},\textbf{V}}+\left[F,\left[G,E\right]^{\textbf{V}}_{\textbf{V},\textbf{A}^{\ast}}\right]^{\textbf{V}}_{\textbf{A}^{\ast},\textbf{V}}\bigg]\nonumber\\&
-d_{i}\bigg[\left[E,\left[F,G\right]^{\textbf{A}^{\ast}}_{\textbf{V},\textbf{A}^{\ast}}\right]^{\textbf{A}^{\ast}}_{\textbf{A}^{\ast},\textbf{A}^{\ast}}+\left[G,\left[E,F\right]^{\textbf{A}^{\ast}}_{\textbf{A}^{\ast},\textbf{V}}\right]^{\textbf{A}^{\ast}}_{\textbf{A}^{\ast},\textbf{A}^{\ast}}+\left[F,\left[G,E\right]^{\textbf{A}^{\ast}}_{\textbf{A}^{\ast},\textbf{A}^{\ast}}\right]^{\textbf{A}^{\ast}}_{\textbf{V},\textbf{A}^{\ast}}\bigg]\nonumber\\&
+d^{2}_{e}\bigg[\left[E,\left[F,G\right]^{\textbf{V}}_{\textbf{V},\textbf{A}^{\ast}}\right]^{\textbf{V}}_{\textbf{A}^{\ast},\textbf{V}}+\left[G,\left[E,F\right]^{\textbf{V}}_{\textbf{A}^{\ast},\textbf{V}}\right]^{\textbf{V}}_{\textbf{A}^{\ast},\textbf{V}}+\left[F,\left[G,E\right]^{\textbf{V}}_{\textbf{A}^{\ast},\textbf{A}^{\ast}}\right]^{\textbf{V}}_{\textbf{V},\textbf{V}}\bigg]\nonumber\\&
+d^{2}_{i}\bigg[\left[E,\left[F,G\right]^{\textbf{A}^{\ast}}_{\textbf{A}^{\ast},\textbf{A}^{\ast}}\right]^{\textbf{A}^{\ast}}_{\textbf{A}^{\ast},\textbf{A}^{\ast}}+\left[G,\left[E,F\right]^{\textbf{A}^{\ast}}_{\textbf{A}^{\ast},\textbf{A}^{\ast}}\right]^{\textbf{A}^{\ast}}_{\textbf{A}^{\ast},\textbf{A}^{\ast}}+\left[F,\left[G,E\right]^{\textbf{A}^{\ast}}_{\textbf{A}^{\ast},\textbf{A}^{\ast}}\right]^{\textbf{A}^{\ast}}_{\textbf{A}^{\ast},\textbf{A}^{\ast}}\bigg]\nonumber\\&
+d^{2}_{e}\bigg[\left[E,\left[F,G\right]^{\textbf{A}^{\ast}}_{\textbf{A}^{\ast},\textbf{A}^{\ast}}\right]^{\textbf{A}^{\ast}}_{\textbf{A}^{\ast},\textbf{A}^{\ast}}+\left[G,\left[E,F\right]^{\textbf{A}^{\ast}}_{\textbf{A}^{\ast},\textbf{A}^{\ast}}\right]^{\textbf{A}^{\ast}}_{\textbf{A}^{\ast},\textbf{A}^{\ast}}+\left[F,\left[G,E\right]^{\textbf{A}^{\ast}}_{\textbf{A}^{\ast},\textbf{A}^{\ast}}\right]^{\textbf{A}^{\ast}}_{\textbf{A}^{\ast},\textbf{A}^{\ast}}\bigg]\nonumber\\&
-d_{i}d^{2}_{e}\bigg[\left[E,\left[F,G\right]^{\textbf{V}}_{\textbf{A}^{\ast},\textbf{A}^{\ast}}\right]^{\textbf{V}}_{\textbf{A}^{\ast},\textbf{V}}+\left[G,\left[E,F\right]^{\textbf{V}}_{\textbf{A}^{\ast},\textbf{A}^{\ast}}\right]^{\textbf{V}}_{\textbf{A}^{\ast},\textbf{V}}+\left[F,\left[G,E\right]^{\textbf{V}}_{\textbf{A}^{\ast},\textbf{A}^{\ast}}\right]^{\textbf{V}}_{\textbf{A}^{\ast},\textbf{V}}\bigg]
\nonumber\\& ~~~+ O(\partial^2).
\nonumber\\
\end{eqnarray}
\normalsize
By lemma\ref{lemma:base_algebra}, only $O(\partial^2)$ terms remain on the right-hand side vanishes.
As we have mentioned, on the other hand, $O(\partial^2)$ vanishes in
$\left\{E,\left\{F,G\right\}\right\}+\circlearrowleft$.
Hence, Jacobi's identity has been proved.

\subsection{Jacobi's identity for Hall MHD system}
\indent \indent The Hamiltonian formulation of Hall MHD is already known\cite{Holm87, Kawazura2012, YoshidaHameiri13}. To see the relation to the extended MHD, we present the formulation.\\ 
Setting the electron skin depth $d_{e}=0$ in the extended MHD model,we obtain the normalized Hall MHD equations,
\begin{eqnarray}
\frac{\partial \rho}{\partial t}=-\nabla\cdot\left(\rho\textbf{V}\right),
\end{eqnarray}
\begin{eqnarray}
\frac{\partial \textbf{V}}{\partial t}=-\left(\nabla\times\textbf{V}\right)\times\textbf{V}-\nabla\left(h+\frac{V^{2}}{2}\right)+\rho^{-1} \left(\nabla\times\textbf{B}\right)\times\textbf{B},
\end{eqnarray},
\begin{eqnarray}
\frac{\partial \textbf{B}}{\partial t}=\nabla\times\left(\textbf{V}\times\textbf{B}\right)- d_{i} \nabla\times\left(\rho^{-1} \left(\nabla\times\textbf{B}\right)\times\textbf{B}\right).
\end{eqnarray}
the state vector is $u=\left(\rho, \textbf{V},\textbf{B}\right)^{t}$. The energy of the Hall MHD reduced into 
\begin{eqnarray}
\mathscr{H}:=\int_{\Omega}\left\{\rho\left(\frac{V^{2}}{2}+ U\left(\rho\right)\right)+\frac{B^{2}}{2}\right\} d^{3}x.
\end{eqnarray}
Also, under the same condition Poisson operator becomes
\begin{eqnarray}
\mathcal{J}_{Hall}= \left(
\begin{array}{ccc}
	0&-\nabla\cdot&0\\
	\\
	-\nabla&-\rho^{-1}\left(\nabla\times\textbf{V}\right)\times\circ&\rho^{-1}\left(\nabla\times\circ\right)\times\textbf{B}\\
	\\
0&\nabla\times\left(\circ\times\rho^{-1}\textbf{B}\right)&-d_{i}\nabla\times\left(\rho^{-1}\left(\nabla\times\circ\right)\times\textbf{B}\right)
\end{array}
\right).
\end{eqnarray}
\subsubsection{Poisson bracket and Jacobi's identity for Hall MHD}
The noncanonical Poisson bracket of the Hall MHD system is given as
\begin{eqnarray}
\left\{F,G\right\}= -\int_{\Omega}&\Big\{\left[F_{\rho} \nabla\cdot G_{\textbf{V}}+F_{\textbf{V}}\cdot\nabla G_{\rho}\right]+\left[F_{\textbf{V}}\cdot\rho^{-1}\left(\left(\nabla\times\textbf{V}\right)\times G_{\textbf{V}}\right)\right]\nonumber \\ 
&-\left[\textbf{B}\cdot\rho^{-1}\left(F_{\textbf{V}}\times \left(\nabla\times G_{\textbf{B}}\right)\right)+\textbf{B}\cdot\rho^{-1}\left(\left(\nabla\times F_{\textbf{B}}\right)\times G_{\textbf{V}}\right)\right]\nonumber \\& 
+ d_{i} \left[ \textbf{B}\cdot\rho^{-1}\left(\left(\nabla\times F_{\textbf{B}}\right)\times \left(\nabla\times G_{\textbf{B}}\right)\right)\right]\Big\}d^{3}x.
\end{eqnarray}
We obtain
\small
\begin{eqnarray}
\fl\left\{E,\left\{F,G\right\}\right\}+\circlearrowleft
=&\int_{\Omega} \big\{\left(\nabla\times\textbf{V}\right)\cdot\big[\rho^{-1}E_{\textbf{V}}\times\big(\nabla\times\rho^{-1}\left(F_{\textbf{V}}\times G_{\textbf{V}}\right)\big)\big]+E_{\textbf{V}}\cdot \big[\nabla \big(\rho^{-2}\left(\nabla\times\textbf{V}\right)\cdot\left(F_{\textbf{V}}\times G_{\textbf{V}}\right)\big)\big]\nonumber\\&  
+\left(\nabla\times\textbf{V}\right)\cdot\big[\rho^{-1}G_{\textbf{V}}\times\big(\nabla\times\rho^{-1}\left(E_{\textbf{V}}\times F_{\textbf{V}}\right)\big)\big]+G_{\textbf{V}}\cdot \big[\nabla \big(\rho^{-2}\left(\nabla\times\textbf{V}\right)\cdot\left(E_{\textbf{V}}\times F_{\textbf{V}}\right)\big)\big]\nonumber\\&
+ \left(\nabla\times\textbf{V}\right)\cdot\big[\rho^{-1}F_{\textbf{V}}\times\big(\nabla\times\rho^{-1}\left(G_{\textbf{V}}\times E_{\textbf{V}}\right)\big)\big]+F_{\textbf{V}}\cdot \big[\nabla \big(\rho^{-2}\left(\nabla\times\textbf{V}\right)\cdot\left(G_{\textbf{V}}\times E_{\textbf{V}}\right)\big)\big]\big\}d^{3}x\nonumber\\&
+\int_{\Omega} \big\{\textbf{B}\cdot\big[\rho^{-1}\left(\nabla\times E_{\textbf{B}}\right)\times\big(\nabla\times\rho^{-1}\left[  F_{\textbf{V}}\times G_{\textbf{V}}\right]\big)\big]\nonumber\\&  
+ \textbf{B}\cdot\big[\rho^{-1}G_{\textbf{V}}\times\big(\nabla\times\rho^{-1}\left[\left(\nabla\times E_{\textbf{B}}\right)\times F_{\textbf{V}}\right]\big)\big]+G_{\textbf{V}}\cdot \nabla \big[\rho^{-2}\textbf{B}\cdot \big( \left(\nabla\times E_{\textbf{B}}\right)\times F_{\textbf{V}}\big)\big]\nonumber\\& 
+\textbf{B}\cdot\big[\rho^{-1}F_{\textbf{V}}\times\big(\nabla\times\rho^{-1}\left[G_{\textbf{V}}\times \left(\nabla\times E_{\textbf{B}}\right)\right]\big)\big]+F_{\textbf{V}}\cdot \nabla \big[\rho^{-2}\textbf{B}\cdot \big( G_{\textbf{V}}\times\left(\nabla\times E_{\textbf{B}}\right)\big)\big]
\big\}d^{3}x\nonumber\\&
+\int_{\Omega} \big\{\textbf{B}\cdot\big[\rho^{-1}E_{\textbf{V}}\times\big(\nabla\times\rho^{-1}\left[F_{\textbf{V}}\times \left(\nabla\times G_{\textbf{B}}\right)\right]\big)\big]+E_{\textbf{V}}\cdot \nabla \big[\rho^{-2}\textbf{B}\cdot \big(F_{\textbf{V}}\times \left(\nabla\times G_{\textbf{B}}\right)\big)\big]\nonumber\\&  
+\textbf{B}\cdot\big[\rho^{-1}\left(\nabla\times G_{\textbf{B}}\right)\times\big(\nabla\times\rho^{-1}\left[E_{\textbf{V}}\times G_{\textbf{V}}\right]\big)\big]\nonumber\\&
+ \textbf{B}\cdot\big[\rho^{-1}F_{\textbf{V}}\times\big(\nabla\times\rho^{-1}\left[\left(\nabla\times G_{\textbf{B}}\right)\times E_{\textbf{V}}\right]\big)\big]+F_{\textbf{V}}\cdot \nabla \big[\rho^{-2}\textbf{B}\cdot \big(\left(\nabla\times G_{\textbf{B}}\right)\times E_{\textbf{V}}\big)\big] \big\}d^{3}x
\nonumber\\&
+\int_{\Omega} \big\{\textbf{B}\cdot\big[\rho^{-1}E_{\textbf{V}}\times\big(\nabla\times\rho^{-1}\left[ \left(\nabla\times F_{\textbf{B}}\right)\times G_{\textbf{V}}\right]\big)\big]+E_{\textbf{V}}\cdot \nabla \big[\rho^{-2}\textbf{B}\cdot \big(\left(\nabla\times F_{\textbf{B}}\right)\times G_{\textbf{V}}\big)\big]\nonumber\\&  
+ \textbf{B}\cdot\big[\rho^{-1}G_{\textbf{V}}\times\big(\nabla\times\rho^{-1}\left[ E_{\textbf{V}}\times\left(\nabla\times F_{\textbf{B}}\right)\right]\big)\big]+G_{\textbf{V}}\cdot \nabla \big[\rho^{-2}\textbf{B}\cdot \big( E_{\textbf{V}}\times \left(\nabla\times F_{\textbf{B}}\right)\big)\big]\nonumber\\& 
+\textbf{B}\cdot\big[\rho^{-1}\left(\nabla\times F_{\textbf{B}}\right)\times\big(\nabla\times\rho^{-1}\left[G_{\textbf{V}}\times E_{\textbf{V}}\right]\big)\big]
\big\}d^{3}x\nonumber\\&
-d_{i}\int_{\Omega} \big\{\textbf{B}\cdot\big[\rho^{-1}E_{\textbf{V}}\times\big(\nabla\times\rho^{-1}\left[ \left(\nabla\times F_{\textbf{B}}\right)\times \left(\nabla\times G_{\textbf{B}}\right)\right]\big)\big]\nonumber\\&+E_{\textbf{V}}\cdot \nabla \big[\rho^{-2}\textbf{B}\cdot \big(\left(\nabla\times F_{\textbf{B}}\right)\times \left(\nabla\times G_{\textbf{B}}\right)\big)\big]\nonumber\\&  
+\textbf{B}\cdot \big[\rho^{-1}\left(\nabla\times G_{\textbf{B}}\right)\times \big(\nabla\times \rho^{-1}\left[E_{\textbf{V}}\times\left(\nabla\times F_{\textbf{B}}\right) \right]\big)\big]\nonumber\\&+\textbf{B}^{\ast}\cdot \big[\rho^{-1}\left(\nabla\times F_{\textbf{B}}\right)\times \big(\nabla\times \rho^{-1}\left[\left(\nabla\times F_{\textbf{B}}\right)\times E_{\textbf{V}} \right]\big)\big]\big\}d^{3}x\nonumber\\&
-d_{i}\int_{\Omega} \big\{\textbf{B}\cdot\big[\rho^{-1}\left(\nabla\times E_{\textbf{B}}\right)\times\big(\nabla\times\rho^{-1}\left[\left(\nabla\times F_{\textbf{B}}\right)\times G_{\textbf{V}}\right]\big)\big] \nonumber\\&
+\textbf{B}\cdot \big[\rho^{-1}G_{\textbf{V}}\times \big(\nabla\times \rho^{-1}\left[\left(\nabla\times E_{\textbf{B}}\right)\times \left(\nabla\times F_{\textbf{B}}\right) \right]\big)\big]\nonumber\\& +G_{\textbf{V}}\cdot \nabla \big[\rho^{-2}\textbf{B}\cdot \big(\left(\nabla\times E_{\textbf{B}}\right)\times \left(\nabla\times F_{\textbf{B}}\right)\big)\big]
\nonumber\\&+\textbf{B}\cdot \big[\rho^{-1}\left(\nabla\times F_{\textbf{B}}\right)\times \big(\nabla\times \rho^{-1}\left[G_{\textbf{V}}\times \left(\nabla\times E_{\textbf{B}}\right) \right]\big)\big]\big\}d^{3}x\nonumber\\&
-d_{i}\int_{\Omega} \big\{\textbf{B}\cdot\big[\rho^{-1}\left(\nabla\times E_{\textbf{B}}\right)\times\big(\nabla\times\rho^{-1}\left[F_{\textbf{V}}\times \left(\nabla\times G_{\textbf{B}}\right)\right]\big)\big]\nonumber\\&
+\textbf{B}\cdot \big[\rho^{-1}\left(\nabla\times G_{\textbf{B}}\right)\times \big(\nabla\times \rho^{-1}\left[\left(\nabla\times E_{\textbf{B}}\right)\times F_{\textbf{V}} \right]\big)\big]\nonumber\\& 
+\textbf{B}\cdot \big[\rho^{-1}F_{\textbf{V}}\times \big(\nabla\times \rho^{-1}\left[\left(\nabla\times G_{\textbf{B}}\right)\times \left(\nabla\times E_{\textbf{B}}\right) \right]\big)\big]\nonumber\\&+ F_{\textbf{V}}\cdot \nabla \big[\rho^{-2}\textbf{B}\cdot \big(\left(\nabla\times G_{\textbf{B}}\right)\times\left(\nabla\times E_{\textbf{B}}\right)\big)\big]\big\}d^{3}x\nonumber\\&
+d^{2}_{i}\int_{\Omega} \big\{\textbf{B}\cdot\big[\rho^{-1}\left(\nabla\times E_{\textbf{B}}\right)\times\big(\nabla\times\rho^{-1}\left[\left(\nabla\times F_{\textbf{B}}\right)\times \left(\nabla\times G_{\textbf{B}}\right)\right]\big)\big]\nonumber\\&+\textbf{B}\cdot\big[\rho^{-1}\left(\nabla\times G_{\textbf{B}}\right)\times\big(\nabla\times\rho^{-1}\left[\left(\nabla\times E_{\textbf{B}}\right)\times \left(\nabla\times F_{\textbf{B}}\right)\right]\big)\big]\nonumber\\&+\textbf{B}\cdot\big[\rho^{-1}\left(\nabla\times F_{\textbf{B}}\right)\times\big(\nabla\times\rho^{-1}\left[\left(\nabla\times G_{\textbf{B}}\right)\times \left(\nabla\times E_{\textbf{B}}\right)\right]\big)\big]\big\}d^{3}x
\nonumber\\& ~~~+O(\partial^2).
\label{eq44}
\end{eqnarray}
\normalsize
We can show that $O(\partial^2)$ terms in (\ref{eq44}) vanishes.
The Poisson bracket of Hall MHD is rewritten as
\begin{eqnarray}
\fl\left\{F,G\right\}= \left[F,G\right]^{\textbf{V}}_{\textbf{V},\textbf{V}}+ \left[F,G\right]^{\textbf{A}}_{\textbf{V},\textbf{A}}+\left[F,G\right]^{\textbf{A}}_{\textbf{A},\textbf{V}}-d_{i}\left[F,G\right]^{\textbf{A}}_{\textbf{A},\textbf{A}}+\int_{\Omega}\left[F_{\rho} \nabla\cdot G_{\textbf{V}}+F_{\textbf{V}}\cdot\nabla G_{\rho}\right]d^{3}x .
\end{eqnarray}
We may write
\begin{eqnarray}
\fl\left\{E,\left\{F,G\right\}\right\}+\circlearrowleft&=\left[E,\left\{F,G\right]^{\textbf{V}}_{\textbf{V},\textbf{V}}\right]^{\textbf{V}}_{\textbf{V},\textbf{V}}+\left[G,\left[E,F\right]^{\textbf{V}}_{\textbf{V},\textbf{V}}\right]^{\textbf{V}}_{\textbf{V},\textbf{V}}+\left[F,\left[G,E\right]^{\textbf{V}}_{\textbf{V},\textbf{V}}\right]^{\textbf{V}}_{\textbf{V},\textbf{V}}\nonumber\\&
+\left[E,\left[F,G\right]^{\textbf{A}}_{\textbf{V},\textbf{V}}\right]^{\textbf{A}}_{\textbf{A},\textbf{A}}+\left[G,\left[E,F\right]^{\textbf{A}}_{\textbf{A},\textbf{V}}\right]^{\textbf{A}}_{\textbf{V},\textbf{A}}+\left[F,\left[G,E\right]^{\textbf{A}}_{\textbf{V},\textbf{A}}\right]^{\textbf{A}}_{\textbf{V},\textbf{A}}\nonumber\\&
+\left[E,\left[F,G\right]^{\textbf{A}}_{\textbf{V},\textbf{A}}\right]^{\textbf{A}}_{\textbf{V},\textbf{A}}+\left[G,\left[E,F\right]^{\textbf{A}}_{\textbf{V},\textbf{V}}\right]^{\textbf{A}}_{\textbf{A},\textbf{A}}+\left[F,\left[G,E\right]^{\textbf{A}}_{\textbf{A},\textbf{V}}\right]^{\textbf{A}}_{\textbf{V},\textbf{A}}\nonumber\\&
+\left[E,\left[F,G\right]^{\textbf{A}}_{\textbf{A},\textbf{V}}\right]^{\textbf{A}}_{\textbf{V},\textbf{A}}+\left[G,\left[E,F\right]^{\textbf{A}}_{\textbf{V},\textbf{A}}\right]^{\textbf{A}}_{\textbf{V},\textbf{A}}+\left[F,\left[G,E\right]^{\textbf{A}}_{\textbf{V},\textbf{V}}\right]^{\textbf{A}}_{\textbf{A},\textbf{A}}\nonumber\\&
-d_{i}\bigg[\left[E,\left[F,G\right]^{\textbf{A}}_{\textbf{A},\textbf{A}}\right]^{\textbf{A}}_{\textbf{V},\textbf{A}}+\left[G,\left[E,F\right]^{\textbf{A}}_{\textbf{V},\textbf{A}}\right]^{\textbf{A}}_{\textbf{A},\textbf{A}}+\left[F,\left[G,E\right]^{\textbf{A}}_{\textbf{A},\textbf{V}}\right]^{\textbf{A}}_{\textbf{A},\textbf{A}}\bigg]\nonumber\\&
-d_{i}\bigg[\left[E,\left[F,G\right]^{\textbf{A}}_{\textbf{A},\textbf{V}}\right]^{\textbf{A}}_{\textbf{A},\textbf{A}}+\left[G,\left[E,F\right]^{\textbf{A}}_{\textbf{A},\textbf{A}}\right]^{\textbf{A}}_{\textbf{V},\textbf{A}}+\left[F,\left[G,E\right]^{\textbf{A}}_{\textbf{V},\textbf{A}}\right]^{\textbf{A}}_{\textbf{A},\textbf{A}}\bigg]\nonumber\\&
-d_{i}\bigg[\left[E,\left[F,G\right]^{\textbf{A}}_{\textbf{V},\textbf{A}}\right]^{\textbf{A}}_{\textbf{A},\textbf{A}}+\left[G,\left[E,F\right]^{\textbf{A}}_{\textbf{A},\textbf{V}}\right]^{\textbf{A}}_{\textbf{A},\textbf{A}}+\left[F,\left[G,E\right]^{\textbf{A}}_{\textbf{A},\textbf{A}}\right]^{\textbf{A}}_{\textbf{V},\textbf{A}}\bigg]\nonumber\\&
+d^{2}_{i}\bigg[\left[E,\left[F,G\right]^{\textbf{A}}_{\textbf{A},\textbf{A}}\right]^{\textbf{A}}_{\textbf{A},\textbf{A}}+\left[G,\left[E,F\right]^{\textbf{A}}_{\textbf{A},\textbf{A}}\right]^{\textbf{A}}_{\textbf{A},\textbf{A}}+\left[F,\left[G,E\right]^{\textbf{A}}_{\textbf{A},\textbf{A}}\right]^{\textbf{A}}_{\textbf{A},\textbf{A}}\bigg]
\nonumber\\& +O(\partial^2),
\end{eqnarray}
which, by Lemma\ref{lemma:base_algebra}, vanishes, proving Jacobi's identity. 

\subsection{Jacobi's identity for inertial MHD system}
\indent \indent The inertial MHD model is obtained by setting the ion skin depth $d_{i}=0$ in the extended MHD model:
\begin{eqnarray}
\frac{\partial \rho}{\partial t}=-\nabla\cdot\left(\rho\textbf{V}\right),
\end{eqnarray}
\begin{eqnarray}
\frac{\partial \textbf{V}}{\partial t}=-\left(\nabla\times\textbf{V}\right)\times\textbf{V}-\nabla\left(h+\frac{V^{2}}{2}\right)+\rho^{-1} \left(\nabla\times\textbf{B}\right)\times\textbf{B}^{\ast}-d^{2}_{e}\nabla\left(\frac{\left(\nabla\times\textbf{B}\right)^{2}}{2\rho^{2}}\right),
\end{eqnarray}
\begin{eqnarray}
\frac{\partial \textbf{B}^{\ast}}{\partial t}=\nabla\times\left(\textbf{V}\times\textbf{B}^{\ast}\right)+d^{2}_{e} \nabla\times\left(\rho^{-1} \left(\nabla\times\textbf{B}\right)\times\left(\nabla\times\textbf{V}\right)\right).
\end{eqnarray}
The energy is\cite{K.Kimura} 
\begin{eqnarray}
\mathscr{H}:=\int_{\Omega}\left\{\rho\left(\frac{V^{2}}{2}+U\left(\rho\right)\right)+\frac{B^{2}}{2}+d^{2}_{e}\frac{\left(\nabla\times\textbf{B}\right)^{2}}{2 \rho}\right\} d^{3}x.
\end{eqnarray}
With respect to the state vector $u=\left(\rho, \textbf{V},\textbf{B}^{\ast}\right)^{t}$,
the Poisson operator of the inertial MHD is
\begin{eqnarray}
\label{50}
\mathcal{J}_{inertial}= \left(
\begin{array}{ccc}
	0&-\nabla\cdot&0\\
	\\
	-\nabla&-\rho^{-1}\left(\nabla\times\textbf{V}\right)\times\circ&\rho^{-1}\left(\nabla\times\circ\right)\times\textbf{B}^{\ast}\\
	\\
0&\nabla\times\left(\circ\times\rho^{-1}\textbf{B}^{\ast}\right)& d^{2}_{e}\nabla\times\left(\rho^{-1}\left(\nabla\times\circ\right)\times\left(\nabla\times\textbf{V}\right)\right)	
\end{array}
\right) .
\end{eqnarray}

\subsubsection{Poisson bracket and Jacobi's identity for inertial MHD}
The Poisson bracket of the inertial MHD system is written as
\begin{eqnarray}
\left\{F,G\right\}= -\int_{\Omega}&\Bigg\{\left[F_{\rho} \nabla\cdot G_{\textbf{V}}+F_{\textbf{V}}\cdot\nabla G_{\rho}\right]-\left[\rho^{-1}\left(\nabla\times\textbf{V}\right)\cdot\big(F_{\textbf{V}}\times G_{\textbf{V}}\big)\right]\nonumber \\ 
&-\left[\textbf{B}^{\ast}\cdot\rho^{-1}\big(F_{\textbf{V}}\times\left(\nabla\times G_{\textbf{B}^{\ast}}\right)\big)+\textbf{B}^{\ast}\cdot\rho^{-1}\big(\left(\nabla\times F_{\textbf{B}^{\ast}}\right)\times G_{\textbf{V}}\big)\right]\nonumber \\&
-d^{2}_{e}\left[ \left(\nabla\times\textbf{V}\right)\cdot\rho^{-1}\big(\left(\nabla\times F_{\textbf{B}^{\ast}}\right)\times\left(\nabla\times G_{\textbf{B}^{\ast}}\right)\big)\right]\Bigg\}d^{3}x .
\end{eqnarray}
We observe
\small
\begin{eqnarray}
\fl\left\{E,\left\{F,G\right\}\right\}+\circlearrowleft&
=\int_{\Omega} \big\{\left(\nabla\times\textbf{V}\right)\cdot\big[\rho^{-1}E_{\textbf{V}}\times\big(\nabla\times\rho^{-1}\left(F_{\textbf{V}}\times G_{\textbf{V}}\right)\big)\big]+E_{\textbf{V}}\cdot \big[\nabla \big(\rho^{-2}\left(\nabla\times\textbf{V}\right)\cdot\left(F_{\textbf{V}}\times G_{\textbf{V}}\right)\big)\big]\nonumber\\&  
+\left(\nabla\times\textbf{V}\right)\cdot\big[\rho^{-1}G_{\textbf{V}}\times\big(\nabla\times\rho^{-1}\left(E_{\textbf{V}}\times F_{\textbf{V}}\right)\big)\big]+G_{\textbf{V}}\cdot \big[\nabla \big(\rho^{-2}\left(\nabla\times\textbf{V}\right)\cdot\left(E_{\textbf{V}}\times F_{\textbf{V}}\right)\big)\big]\nonumber\\&
+ \left(\nabla\times\textbf{V}\right)\cdot\big[\rho^{-1}F_{\textbf{V}}\times\big(\nabla\times\rho^{-1}\left(G_{\textbf{V}}\times E_{\textbf{V}}\right)\big)\big]+F_{\textbf{V}}\cdot \big[\nabla \big(\rho^{-2}\left(\nabla\times\textbf{V}\right)\cdot\left(G_{\textbf{V}}\times E_{\textbf{V}}\right)\big)\big]\big\}d^{3}x\nonumber\\&
+\int_{\Omega} \big\{\textbf{B}^{\ast}\cdot\big[\rho^{-1}\left(\nabla\times E_{\textbf{B}^{\ast}}\right)\times\big(\nabla\times\rho^{-1}\left[  F_{\textbf{V}}\times G_{\textbf{V}}\right]\big)\big]\nonumber\\&  
+ \textbf{B}^{\ast}\cdot\big[\rho^{-1}G_{\textbf{V}}\times\big(\nabla\times\rho^{-1}\left[\left(\nabla\times E_{\textbf{B}^{\ast}}\right)\times F_{\textbf{V}}\right]\big)\big]+G_{\textbf{V}}\cdot \nabla \big[\rho^{-2}\textbf{B}^{\ast}\cdot \big( \left(\nabla\times E_{\textbf{B}^{\ast}}\right)\times F_{\textbf{V}}\big)\big]\nonumber\\& 
+\textbf{B}^{\ast}\cdot\big[\rho^{-1}F_{\textbf{V}}\times\big(\nabla\times\rho^{-1}\left[G_{\textbf{V}}\times \left(\nabla\times E_{\textbf{B}^{\ast}}\right)\right]\big)\big]+F_{\textbf{V}}\cdot \nabla \big[\rho^{-2}\textbf{B}^{\ast}\cdot \big( G_{\textbf{V}}\times\left(\nabla\times E_{\textbf{B}^{\ast}}\right)\big)\big]
\big\}d^{3}x\nonumber\\&
+\int_{\Omega} \big\{\textbf{B}^{\ast}\cdot\big[\rho^{-1}E_{\textbf{V}}\times\big(\nabla\times\rho^{-1}\left[F_{\textbf{V}}\times \left(\nabla\times G_{\textbf{B}^{\ast}}\right)\right]\big)\big]+E_{\textbf{V}}\cdot \nabla \big[\rho^{-2}\textbf{B}^{\ast}\cdot \big(F_{\textbf{V}}\times \left(\nabla\times G_{\textbf{B}^{\ast}}\right)\big)\big]\nonumber\\&  
+\textbf{B}^{\ast}\cdot\big[\rho^{-1}\left(\nabla\times G_{\textbf{B}^{\ast}}\right)\times\big(\nabla\times\rho^{-1}\left[E_{\textbf{V}}\times G_{\textbf{V}}\right]\big)\big]\nonumber\\&
+ \textbf{B}^{\ast}\cdot\big[\rho^{-1}F_{\textbf{V}}\times\big(\nabla\times\rho^{-1}\left[\left(\nabla\times G_{\textbf{B}^{\ast}}\right)\times E_{\textbf{V}}\right]\big)\big]+F_{\textbf{V}}\cdot \nabla \big[\rho^{-2}\textbf{B}^{\ast}\cdot \big(\left(\nabla\times G_{\textbf{B}^{\ast}}\right)\times E_{\textbf{V}}\big)\big] \big\}d^{3}x
\nonumber\\&
+\int_{\Omega} \big\{\textbf{B}^{\ast}\cdot\big[\rho^{-1}E_{\textbf{V}}\times\big(\nabla\times\rho^{-1}\left[ \left(\nabla\times F_{\textbf{B}^{\ast}}\right)\times G_{\textbf{V}}\right]\big)\big]+E_{\textbf{V}}\cdot \nabla \big[\rho^{-2}\textbf{B}^{\ast}\cdot \big(\left(\nabla\times F_{\textbf{B}^{\ast}}\right)\times G_{\textbf{V}}\big)\big]\nonumber\\&  
+ \textbf{B}^{\ast}\cdot\big[\rho^{-1}G_{\textbf{V}}\times\big(\nabla\times\rho^{-1}\left[ E_{\textbf{V}}\times\left(\nabla\times F_{\textbf{B}^{\ast}}\right)\right]\big)\big]+G_{\textbf{V}}\cdot \nabla \big[\rho^{-2}\textbf{B}^{\ast}\cdot \big( E_{\textbf{V}}\times \left(\nabla\times F_{\textbf{B}^{\ast}}\right)\big)\big]\nonumber\\& 
+\textbf{B}^{\ast}\cdot\big[\rho^{-1}\left(\nabla\times F_{\textbf{B}^{\ast}}\right)\times\big(\nabla\times\rho^{-1}\left[G_{\textbf{V}}\times E_{\textbf{V}}\right]\big)\big]
\big\}d^{3}x\nonumber\\&
+d^{2}_{e}\int_{\Omega} \big\{\left(\nabla\times \textbf{V}\right)\cdot\big[\rho^{-1}E_{\textbf{V}}\times\big(\nabla\times\rho^{-1}\left[ \left(\nabla\times F_{\textbf{B}^{\ast}}\right)\times \left(\nabla\times G_{\textbf{B}^{\ast}}\right)\right]\big)\big]\nonumber\\&+E_{\textbf{V}}\cdot \nabla \big[\rho^{-2}\left(\nabla\times \textbf{V}\right)\cdot \big(\left(\nabla\times F_{\textbf{B}^{\ast}}\right)\times \left(\nabla\times G_{\textbf{B}^{\ast}}\right)\big)\big]\nonumber\\&  
+\left(\nabla\times \textbf{V}\right)\cdot \big[\rho^{-1}\left(\nabla\times G_{\textbf{B}^{\ast}}\right)\times \big(\nabla\times \rho^{-1}\left[E_{\textbf{V}}\times\left(\nabla\times F_{\textbf{B}^{\ast}}\right) \right]\big)\big]\nonumber\\&+\left(\nabla\times \textbf{V}\right)\cdot \big[\rho^{-1}\left(\nabla\times F_{\textbf{B}^{\ast}}\right)\times \big(\nabla\times \rho^{-1}\left[\left(\nabla\times F_{\textbf{B}^{\ast}}\right)\times E_{\textbf{V}} \right]\big)\big]\big\}d^{3}x\nonumber\\&
+d^{2}_{e}\int_{\Omega} \big\{\left(\nabla\times \textbf{V}\right)\cdot\big[\rho^{-1}\left(\nabla\times E_{\textbf{B}^{\ast}}\right)\times\big(\nabla\times\rho^{-1}\left[\left(\nabla\times F_{\textbf{B}^{\ast}}\right)\times G_{\textbf{V}}\right]\big)\big]\nonumber\\& 
+\left(\nabla\times \textbf{V}\right)\cdot \big[\rho^{-1}G_{\textbf{V}}\times \big(\nabla\times \rho^{-1}\left[\left(\nabla\times E_{\textbf{B}^{\ast}}\right)\times \left(\nabla\times F_{\textbf{B}^{\ast}}\right) \right]\big)\big]\nonumber\\& +G_{\textbf{V}}\cdot \nabla \big[\rho^{-2}\left(\nabla\times \textbf{V}\right)\cdot \big(\left(\nabla\times E_{\textbf{B}^{\ast}}\right)\times \left(\nabla\times F_{\textbf{B}^{\ast}}\right)\big)\big]\nonumber\\&
+\left(\nabla\times \textbf{V}\right)\cdot \big[\rho^{-1}\left(\nabla\times F_{\textbf{B}^{\ast}}\right)\times \big(\nabla\times \rho^{-1}\left[G_{\textbf{V}}\times \left(\nabla\times E_{\textbf{B}^{\ast}}\right) \right]\big)\big]\big\}d^{3}x\nonumber\\&
+d^{2}_{e}\int_{\Omega} \big\{\left(\nabla\times \textbf{V}\right)\cdot\big[\rho^{-1}\left(\nabla\times E_{\textbf{B}^{\ast}}\right)\times\big(\nabla\times\rho^{-1}\left[F_{\textbf{V}}\times \left(\nabla\times G_{\textbf{B}^{\ast}}\right)\right]\big)\big]\nonumber\\&
+\left(\nabla\times \textbf{V}\right)\cdot \big[\rho^{-1}\left(\nabla\times G_{\textbf{B}^{\ast}}\right)\times \big(\nabla\times \rho^{-1}\left[\left(\nabla\times E_{\textbf{B}^{\ast}}\right)\times F_{\textbf{V}} \right]\big)\big]\nonumber\\& 
+\left(\nabla\times \textbf{V}\right)\cdot \big[\rho^{-1}F_{\textbf{V}}\times \big(\nabla\times \rho^{-1}\left[\left(\nabla\times G_{\textbf{B}^{\ast}}\right)\times \left(\nabla\times E_{\textbf{B}^{\ast}}\right) \right]\big)\big]\nonumber\\&+ F_{\textbf{V}}\cdot \nabla \big[\rho^{-2}\left(\nabla\times\textbf{V}\right)\cdot \big(\left(\nabla\times G_{\textbf{B}^{\ast}}\right)\times\left(\nabla\times E_{\textbf{B}^{\ast}}\right)\big)\big]\big\}d^{3}x\nonumber\\&
+d^{2}_{e}\int_{\Omega} \big\{\textbf{B}^{\ast}\cdot\big[\rho^{-1}\left(\nabla\times E_{\textbf{B}^{\ast}}\right)\times\big(\nabla\times\rho^{-1}\left[\left(\nabla\times F_{\textbf{B}^{\ast}}\right)\times \left(\nabla\times G_{\textbf{B}^{\ast}}\right)\right]\big)\big]\nonumber\\&+\textbf{B}^{\ast}\cdot\big[\rho^{-1}\left(\nabla\times G_{\textbf{B}^{\ast}}\right)\times\big(\nabla\times\rho^{-1}\left[\left(\nabla\times E_{\textbf{B}^{\ast}}\right)\times \left(\nabla\times F_{\textbf{B}^{\ast}}\right)\right]\big)\big]\nonumber\\&+\textbf{B}^{\ast}\cdot\big[\rho^{-1}\left(\nabla\times F_{\textbf{B}^{\ast}}\right)\times\big(\nabla\times\rho^{-1}\left[\left(\nabla\times G_{\textbf{B}^{\ast}}\right)\times \left(\nabla\times E_{\textbf{B}^{\ast}}\right)\right]\big)\big]\big\}d^{3}x\nonumber\\
\nonumber\\& ~~~+O(\partial^2).
\end{eqnarray}
\normalsize
The Poisson bracket can be written as
\begin{eqnarray}
\fl\left\{F,G\right\}= \left[F,G\right]^{\textbf{V}}_{\textbf{V},\textbf{V}}+ \left[F,G\right]^{\textbf{A}^{\ast}}_{\textbf{V},\textbf{A}^{\ast}}+\left[F,G\right]^{\textbf{A}^{\ast}}_{\textbf{A}^{\ast},\textbf{V}}+d^{2}_{e}\left[F,G\right]^{\textbf{V}}_{\textbf{A}^{\ast},\textbf{A}^{\ast}}+\int_{\Omega}\left[F_{\rho} \nabla\cdot G_{\textbf{V}}+F_{\textbf{V}}\cdot\nabla G_{\rho}\right]d^{3}x .
\end{eqnarray}
We may write
\begin{eqnarray}
\fl\left\{E,\left\{F,G\right\}\right\}+\circlearrowleft&=\left[E,\left[F,G\right]^{\textbf{V}}_{\textbf{V},\textbf{V}}\right]^{\textbf{V}}_{\textbf{V},\textbf{V}}+\left[G,\left[E,F\right]^{\textbf{V}}_{\textbf{V},\textbf{V}}\right]^{\textbf{V}}_{\textbf{V},\textbf{V}}+\left[F,\left[G,E\right]^{\textbf{V}}_{\textbf{V},\textbf{V}}\right]^{\textbf{V}}_{\textbf{V},\textbf{V}}\nonumber\\&
+\left[E,\left[F,G\right]^{\textbf{A}^{\ast}}_{\textbf{V},\textbf{V}}\right]^{\textbf{A}^{\ast}}_{\textbf{A}^{\ast},\textbf{A}^{\ast}}+\left[G,\left[E,F\right]^{\textbf{A}^{\ast}}_{\textbf{A}^{\ast},\textbf{V}}\right]^{\textbf{A}^{\ast}}_{\textbf{V},\textbf{A}^{\ast}}+\left[F,\left[G,E\right]^{\textbf{A}^{\ast}}_{\textbf{V},\textbf{A}^{\ast}}\right]^{\textbf{A}^{\ast}}_{\textbf{V},\textbf{A}^{\ast}}\nonumber\\&
+\left[E,\left[F,G\right]^{\textbf{A}^{\ast}}_{\textbf{V},\textbf{A}^{\ast}}\right]^{\textbf{A}^{\ast}}_{\textbf{V},\textbf{A}^{\ast}}+\left[G,\left[E,F\right]^{\textbf{A}^{\ast}}_{\textbf{V},\textbf{V}}\right]^{\textbf{A}^{\ast}}_{\textbf{A}^{\ast},\textbf{A}^{\ast}}+\left[F,\left[G,E\right]^{\textbf{A}^{\ast}}_{\textbf{A}^{\ast},\textbf{V}}\right]^{\textbf{A}^{\ast}}_{\textbf{V},\textbf{A}^{\ast}}\nonumber\\&
+\left[E,\left[F,G\right]^{\textbf{A}^{\ast}}_{\textbf{A}^{\ast},\textbf{V}}\right]^{\textbf{A}^{\ast}}_{\textbf{V},\textbf{A}^{\ast}}+\left[G,\left[E,F\right]^{\textbf{A}^{\ast}}_{\textbf{V},\textbf{A}^{\ast}}\right]^{\textbf{A}^{\ast}}_{\textbf{V},\textbf{A}^{\ast}}+\left[F,\left[G,E\right]^{\textbf{A}^{\ast}}_{\textbf{V},\textbf{V}}\right]^{\textbf{A}^{\ast}}_{\textbf{A}^{\ast},\textbf{A}^{\ast}}\nonumber\\&
+d^{2}_{e}\bigg[\left[E,\left[F,G\right]^{\textbf{V}}_{\textbf{A}^{\ast},\textbf{A}^{\ast}}\right]^{\textbf{V}}_{\textbf{V},\textbf{V}}+\left[G,\left[E,F\right]^{\textbf{V}}_{\textbf{V},\textbf{A}^{\ast}}\right]^{\textbf{V}}_{\textbf{A}^{\ast},\textbf{V}}+\left[F,\left[G,E\right]^{\textbf{V}}_{\textbf{A}^{\ast},\textbf{V}}\right]^{\textbf{V}}_{\textbf{A}^{\ast},\textbf{V}}\bigg]\nonumber\\&
+d^{2}_{e}\bigg[\left[E,\left[F,G\right]^{\textbf{V}}_{\textbf{A}^{\ast},\textbf{V}}\right]^{\textbf{V}}_{\textbf{A}^{\ast},\textbf{V}}+\left[G,\left[E,F\right]^{\textbf{V}}_{\textbf{A}^{\ast},\textbf{A}^{\ast}}\right]^{\textbf{V}}_{\textbf{V},\textbf{V}}+\left[F,\left[G,E\right]^{\textbf{V}}_{\textbf{V},\textbf{A}^{\ast}}\right]^{\textbf{V}}_{\textbf{A}^{\ast},\textbf{V}}\bigg]\nonumber\\&
+d^{2}_{e}\bigg[\left[E,\left[F,G\right]^{\textbf{V}}_{\textbf{V},\textbf{A}^{\ast}}\right]^{\textbf{V}}_{\textbf{A}^{\ast},\textbf{V}}+\left[G,\left[E,F\right]^{\textbf{V}}_{\textbf{A}^{\ast},\textbf{V}}\right]^{\textbf{V}}_{\textbf{A}^{\ast},\textbf{V}}+\left[F,\left[G,E\right]^{\textbf{V}}_{\textbf{A}^{\ast},\textbf{A}^{\ast}}\right]^{\textbf{V}}_{\textbf{V},\textbf{V}}\bigg]\nonumber\\&
+d^{2}_{e}\bigg[\left[E,\left[F,G\right]^{\textbf{A}^{\ast}}_{\textbf{A}^{\ast},\textbf{A}^{\ast}}\right]^{\textbf{A}^{\ast}}_{\textbf{A}^{\ast},\textbf{A}^{\ast}}+\left[G,\left[E,F\right]^{\textbf{A}^{\ast}}_{\textbf{A}^{\ast},\textbf{A}^{\ast}}\right]^{\textbf{A}^{\ast}}_{\textbf{A}^{\ast},\textbf{A}^{\ast}}+\left[F,\left[G,E\right]^{\textbf{A}^{\ast}}_{\textbf{A}^{\ast},\textbf{A}^{\ast}}\right]^{\textbf{A}^{\ast}}_{\textbf{A}^{\ast},\textbf{A}^{\ast}}\bigg]
\nonumber\\& ~~~+O(\partial^2).
\nonumber\\&
\end{eqnarray}
As in the previous cases, $O(\partial^2)$ terms cancels.  Hence, by Lemma\ref{lemma:base_algebra}, we obtain
Jacobi's identity.

\section{Conclusions and Remarks}
\indent \indent We have formulated the Hamiltonian and Poisson bracket for the extended MHD
which subsumes the Hall MHD and the inertial MHD systems. 
In proving Jacobi's identity, we have unearthed an underlying algebraic relation
that is represented by a generating bracket (\ref{18}) satisfying an extended permutation law.

The formulated Poisson algebra has a nontrivial center (i.e., the Hamiltonian system is concanonical).
The noncanonicality is a common feature of fluid/plasma systems represented by Eulerian variables.
The metamorphoses of Casimir leaves, in response to the singular perturbations scaled by ion and electron skin depths.
\\
The Poisson bracket of the extended MHD system has three independent Casimir elements:
\begin{eqnarray}
\label{55}
C_{1}=\int_{\Omega}\rho~d^{3}x,
\end{eqnarray}
\begin{eqnarray}
\label{56}
C_{2}=\int_{\Omega}\textbf{B}^{\ast}\cdot\left(\bi{V}-\frac{d_{i}}{d^{2}_{e}}\textbf{A}^{\ast}\right) d^{3}x,
\end{eqnarray}
\begin{eqnarray}
\label{57}
C_{3}=\int_{\Omega}\big[\textbf{B}^{\ast}\cdot\textbf{A}^{\ast}+d^{2}_{e}\bi{V}\cdot\left(\nabla\times\bi{V}\right)\big] d^{3}x.
\end{eqnarray}
Combining $C_{2}$ and $C_{3}$, we may define a ``canonical helicity''
\begin{eqnarray}
\label{58}
C_{2,3}&= 2\lambda C_{2}+\frac{1}{d^{2}_{e}}C_{3} 
\nonumber\\&
=\int_{\Omega}\bi{P}^{\ast}\cdot\left(\nabla\times \bi{P}^{\ast}\right) d^{3}x ,
\end{eqnarray}
where $\bi{P}^{\ast}=\bi{V}+\lambda\textbf{A}^{\ast}$ and $\lambda=\frac{-d_{i}+\sqrt{d^{2}_{i}+4d^{2}_{e}}}{2d^{2}_{e}}$.\\
The inertial MHD system also has three independent Casimir elements:
\begin{eqnarray}
C^{'}_{1}=\int_{\Omega}\rho~d^{3}x,
\end{eqnarray}
\begin{eqnarray}
C^{'}_{2}=\int_{\Omega}\bi{V}\cdot \textbf{B}^{\ast} d^{3}x,
\end{eqnarray}
\begin{eqnarray}
C^{'}_{3}=\int_{\Omega}\big[\textbf{B}^{\ast}\cdot\textbf{A}^{\ast}+d^{2}_{e}\bi{V}\cdot\left(\nabla\times\bi{V}\right)\big] d^{3}x.
\end{eqnarray}
Combining $C'_{2}$and $C'_{3}$, we may define a canonical helicity
\begin{eqnarray}
C^{'}_{2,3}&= \frac{2}{d_{e}} C'_{2}+\frac{1}{d^{2}_{e}}C'_{3} 
\nonumber\\&
=\int_{\Omega}\bi{P}^{'\ast}\cdot\left(\nabla\times \bi{P}^{'\ast}\right) d^{3}x ,
\end{eqnarray}
where $\bi{P}^{'\ast}=\bi{V}+\frac{1}{d_{e}}\textbf{A}^{\ast}$. 
The transition from the generalized MHD to the inertial MHD is, therefore, a smooth limit of
$d_{i} \rightarrow0$.
However, the limit of $d_{e} \rightarrow0$ (i.e., generalized MHD $\rightarrow$ Hall MHD), 
and the limit of $d_{i} \rightarrow0$ under $d_{e} =0$ (i.e., Hall MHD $\rightarrow$ ideal MHD;
see \cite{YoshidaHameiri13}) are not that simple.
These singular perturbations will be discussed elsewhere.

\ack
We acknowledge the discussions and suggestions of Professor P. J. Morrison. We thank Mr. Tanehashi and Mr. Takabashi for fruitful discussion about Casimir elements. H. M. Abdelhamid would like to thank the Egyptian Ministry of Higher Education for supporting his research activities.
This work was partly supported by the Grant-in-Aid for Scientific Research (23224014) from MEXT-Japan.

\end{document}